\documentclass[sigconf, nonacm]{acmart}

\usepackage[utf8]{inputenc}
\usepackage{hyperref}
\usepackage{url}
\usepackage{todonotes}
\usepackage{graphicx}
\usepackage[flushleft]{threeparttable}
\usepackage{tabulary}
\usepackage{tabularx}

\usepackage{multirow}
\usepackage{siunitx}
\usepackage{tablefootnote}
\usepackage{mathtools}
\usepackage{cleveref}

\newcommand\vldbdoi{XX.XX/XXX.XX}

\newcommand\vldbavailabilityurl{https://github.com/edualc/lqo_ml_perspective}

\begin{document}
\title{Is Your Learned Query Optimizer Behaving As You Expect? \\ A Machine Learning Perspective} 

\author{Claude Lehmann}
\email{claude.lehmann@zhaw.ch}
\affiliation{
    \institution{Zurich University of Applied Sciences}
    \city{Winterthur}
    \country{Switzerland}
}
\authornote{Both authors contributed equally to this work.}

\author{Pavel Sulimov}
\email{pavel.sulimov@zhaw.ch}
\affiliation{
    \institution{Zurich University of Applied Sciences}
    \city{Winterthur}
    \country{Switzerland}
}
\authornotemark[1]

\author{Kurt Stockinger}
\email{kurt.stockinger@zhaw.ch}
\affiliation{
    \institution{Zurich University of Applied Sciences}
    \city{Winterthur}
    \country{Switzerland}
}

\date{15 January 2023}

\begin{abstract}
The current boom of learned query optimizers (LQO) can be explained not only by the general continuous improvement of deep learning (DL) methods but also by the straightforward formulation of a query optimization problem (QOP) as a machine learning (ML) one. The idea is often to replace dynamic programming approaches, widespread for solving QOP, with more powerful methods such as reinforcement learning. However, such a rapid "game change" in the field of QOP could not pass without consequences - other parts of the ML pipeline, except for predictive model development, have large improvement potential. For instance, different LQOs introduce their own restrictions on training data generation from queries, use an arbitrary train/validation approach, and evaluate on a voluntary split of benchmark queries. 

In this paper, we attempt to standardize the ML pipeline for evaluating LQOs by introducing a new \emph{end-to-end benchmarking framework}. Additionally, we guide the reader through each data science stage in the ML pipeline and provide novel insights from the machine learning perspective, considering the specifics of QOP. Finally, we perform a \emph{rigorous evaluation of existing LQOs, showing that PostgreSQL outperforms these LQOs in almost all experiments depending on the train/test splits}.

\end{abstract}

\maketitle



\section{Introduction}
\label{introduction}

Over the last decade, machine learning (ML) approaches have heavily dominated classical query optimization methods. Having in total $O(n!)$ possible logical plans in the worst case for queries where the join graph is a clique with $n$ tables, the problem is classified as NP-hard \cite{wang1996complexity}. This implies that exhaustive methods cannot solve the problem for a higher order of joins\footnote{PostgreSQL abandons exhaustive methods for queries with 12 or more \texttt{FROM} items.}, thus demanding the need for heuristical approaches.

\begin{figure}[h!]
    \centering
    \includegraphics[width=\linewidth, keepaspectratio]{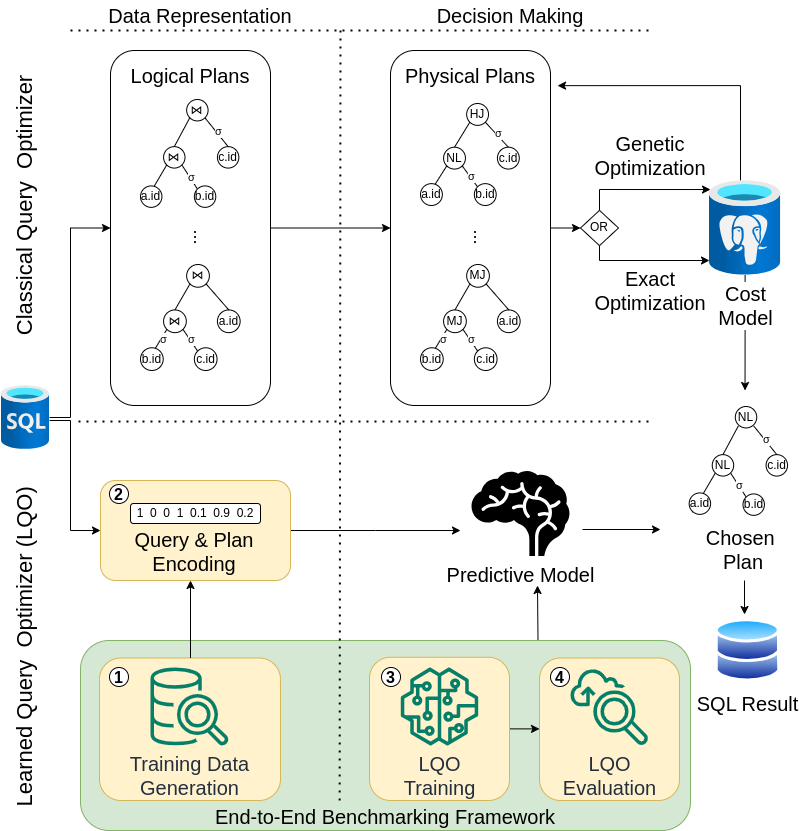}
    \caption{Comparison of classical and learned query optimizers (LQO) - see top and bottom halves, respectively. The stages (1) Training Data Generation, (3) LQO Training, and (4) LQO Evaluation are the primary components of our End-to-End Benchmarking Framework. Together with the (2) Query \& Plan Encoding stage, they form the typical machine learning pipeline for a LQO.}
    \label{fig:mlops_architecture}
\end{figure}

In Figure \ref{fig:mlops_architecture}, we compare typical pipelines for classical and learned query optimizers. The \textit{classical approach}, implemented inside database management systems (DBMS), has the stages of query representation via logical and physical plans, with a follow-up search of an optimal plan using cardinality-based cost model estimations. In addition to dynamic programming-based methods, genetic algorithms~\cite{geqo_pg_doc} are also used since they are proven to be more efficient for queries with a high number of joins~\cite{petkovic2011dp_vs_ga}. 

The bottom part of Figure \ref{fig:mlops_architecture} shows \textit{learned query optimizers} (LQO), the most recent trend for end-to-end query optimization. These approaches require a more complicated pipeline because of the use of ML methods. Looking at it from the \textit{ML perspective}, the pipeline should consist of several stages: (1) training data generation, (2) query \& plan encoding, (3) ML model training, and (4) ML model evaluation. The violation of theoretical ML principles~\cite{russel2010ml} at each stage and the absence of a unified reproducible framework make it currently \emph{impossible to fairly compare the results of LQOs}.

Let us briefly describe what can go wrong at each stage, i.e., the major challenges of the ML pipeline for LQOs from both a data science and an engineering perspective and how we solve them as \underline{contributions of this paper}.

\textbf{Training Data Generation.\footnote{The points about training data also apply to validation and test data.}} \emph{When no ready-to-use training data is provided for benchmarking, opportunities for biased data creation appear}. For LQOs, we observed that only the queries are given as SQL statements for popular benchmarks such as JOB \cite{JOB_leis2015good}. The key problem is that these statements cannot be explicitly used as input for ML models without querying the databases (DB) and extracting metadata such as cardinalities or execution times. This implies a gap between the given benchmark data and the actual features used to train LQOs, which are strongly correlated with the parametric conditions when querying the database. 

\underline{Contribution:} \emph{We discuss general limitations that could hamper the process of similar training data creation 
in Section \ref{sec:train_data_generation}}.

\textbf{Query \& Plan Encoding.} \emph{Encoding the queries such that the principle of invariance\footnote{We formulate the principle of invariance \cite{la1976stability} in data generation as follows: When given the same input, the data generating system should return the same output.} is broken, leads to inconsistencies in the performance.} When different queries are encoded using column selectivities, it is possible that large sections of the encoding (or even the full vectors) are identical. This is because many filter combinations result in the same selectivity. Hence, the model would potentially suffer from the mismatch between features and target variables and will only perform well if this inconsistency is mitigated.

\underline{Contribution}: \emph{We diagnose the invariance issues in particular methods and give encoding recommendations in Section \ref{sec:query_and_plan_encoding}.}

\textbf{LQO Training.} \emph{Contravening the basic training techniques and misapplying mathematical models makes your ML model behave unexpectedly}. Complicated DL models are hard to train, which makes hyperparameter tuning and validation procedures the cornerstone for gaining high predictive power. Moreover, injecting additional mathematical mechanisms can have adverse side effects that negatively impact the training itself and, in turn, the query performance. 

\underline{Contribution:} \emph{We propose enhancements to make the training process of LQO methods more stable and reliable in Section \ref{sec:lqo_training}.}


\textbf{LQO Evaluation.} \emph{When your model is trained and then evaluated on a non-fixed train/test split, comparisons become data-centric rather than model-centric}, i.e., the choice of the train/test split strongly correlates with the model's performance. For example, the performance on two different train/test splits of the same type (such as randomly splitting queries) is not comparable. Explicit examples of this can be seen in Figures \ref{fig:results_detailed__test__JOB} and \ref{fig:results_detailed__test__STACK} of our experiments). Despite the existence of public query optimization benchmarks like the one in \cite{JOB_leis2015good} and \cite{marcus2021bao}, it remains an open question which queries serve as the train data and which ones are the test data. The attempts to suggest the procedure of a unified evaluation were only made recently in \cite{marcus12neo}, \cite{balsa_Yang2022} and \cite{zhang2023adaptive}. 

\underline{Contribution:} \emph{We unify data splitting for LQOs and introduce a procedure to test different levels of generalization in Section \ref{sec:lqo_evaluation}.}

\textbf{Reproducibility.} \emph{Any ML method developed in academia has negligible practical value if it cannot be reproduced on arbitrary software and hardware}. With ML approaches finding widespread use in academic research, navigating the realm of learned query optimization presents challenges, as it requires proficiency in the core subject of database research and numerous related engineering fields. These approaches typically require complex programming code and (ML) models with inherent stochasticity. Hence, reproducibility is becoming a growing concern in academia.

\underline{Contribution:} \emph{We suggest an \textbf{End-to-End Benchmarking Framework} - a novel meta-benchmarking framework that is capable of equalizing the conditions under which the ML-based LQOs are trained and tested, guaranteeing consistency in comparisons in Section \ref{sec:framework}.}

\underline{Main contribution}: We perform an extensive evaluation of existing LQOs using our end-to-end benchmarking framework in Section \ref{sec:experiments}. \textbf{Our results demonstrate that current LQOs do not systematically perform better than PostgreSQL.} These findings indicate that \emph{novel research is required to make LQOs competitive} with more traditional approaches - and not only in specific cases.

The paper is organized as follows: First, we briefly review recent LQO methods in Section \ref{sec:related_work}. Then, we dissect the data science stages in the ML pipeline applied to query optimization, not only discussing potential hurdles that can occur while processing each stage but also suggesting ways of mitigating them via practical (see Sections \ref{sec:train_data_generation} \& \ref{sec:lqo_evaluation}) and theoretical (see Sections \ref{sec:query_and_plan_encoding} \& \ref{sec:lqo_training}) recommendations. Based on the challenges in the reproducibility of ML approaches, we propose our End-to-End Benchmarking Framework in Section \ref{sec:framework}. Afterwards, we perform an elaborate experimental evaluation of recently released LQOs from an ML perspective in Section \ref{sec:experiments}. Finally, we conclude the paper in Section \ref{sec:conclusion}.


\section{Related Work}
\label{sec:related_work}

Before end-to-end LQOs appeared, significant progress had been made toward using modern ML approaches for query optimization. For instance, DQ~\cite{krishnan2018dq}, ReJOIN~\cite{marcus2018rejoin1, marcus2018rejoin2}, and others \cite{krishnan2018learning, heitz2019join} apply reinforcement learning (RL) in an exploration-exploitation strategy with the goal of finding the optimal join order. These methods use a cost model to produce a "join score" reward for the learning agent. 

The first end-to-end LQO Neo~\cite{marcus12neo} uses a neural network (NN) to estimate the latencies of a full query plan given a sub-plan as an input. The optimal plan is predicted via a greedy tree search in the join and scan space and consecutive bottom-up plan construction.

RTOS~\cite{yu2020rtos} assumes that the join graph is built as a sequence of join operations between two tables, ignoring scans, and applies a graph NN to train an RL agent. The predicted query plan is built similarly to Neo, though it applies a depth-first search.

Bao~\cite{marcus2021bao} sits on top of the PostgreSQL query optimizer, controlling the execution flow by enabling or disabling a subset of join and scan operations. These subsets are referred to as hint sets, and Bao provides neither the full join order nor which scan types are used for which table but rather advises which operations not to use. 

Balsa~\cite{balsa_Yang2022} is based on the same architecture as Neo. However, it introduces several modifications to the training pipeline: it pre-trains using the cost model estimations of a DBMS instead of real latencies, it uses timeouts during query executions, and it does not sample training data from the replay buffer but rather uses the data points produced by the most recent NN state.

Lero~\cite{zhu2023lero} formulates the problem as a learning-to-rank (LTR) task and generates various candidate query plans from the DBMS by changing the internal cardinality estimations. The plan comparator module selects the better of two generated candidate plans, similarly choosing the optimal plan during inference.
LEON~\cite{chen2023leon} is another LTR method. Unlike Lero, it brute-forces many possible physical plans in a dynamic programming manner and prunes them before training. Training happens only on the top chosen SQL/query plan-pairs, ranked by their latency and posterior uncertainty estimation obtained from a Bayesian NN.

LOGER~\cite{chen2023loger} uses the conceptual ML model pipeline from RTOS, though extending the action space for join order recommendations by adding the join type. LOGER restricts the operation recommendation, i.e., which join type not to use, by applying $\epsilon$-beam search for plan prediction.

HybridQO~\cite{yu2022hybridqo} uses a mix of cost and latency estimations, like some other methods, but in a different manner: it first gets the candidate plans from the DBMS via hints. Those hints are obtained from the top levels of the query plan tree explored by a Monte-Carlo Tree Search (MCTS) with an upper confidence bound and using the cost as a target (the cost is estimated with an NN from RTOS). Then, the same network architecture is used to predict the latency and uncertainty from the candidate plans. A multi-head performance estimator makes the final plan selection.

In the recent paper~\cite{zhang2023adaptive}, the authors question the reasonability of training complicated and computationally costly LQOs. As an alternative, they suggest the combination of look-ahead information passing (LIP), in which adaptive semi-join techniques and adaptive join algorithms (AJA) are used. The latter checks whether a hash join should be replaced by a nested loop join at runtime.

In this paper, we introduce neither a new LQO nor a classical alternative. Instead, we provide \emph{recommendations to improve LQOs} based on a \emph{vast evaluation of existing LQOs from an ML perspective}.


\section{Training Data Generation}
\label{sec:train_data_generation}

Typically, ML problems have publicly available benchmarks with ready-to-use training data that is identical for all participants. QOP benchmarks differ regarding the provided data and only serve as a source for generating the training data, suitable as an input into ML models. This makes the whole ML pipeline vulnerable to inconsistencies in the data generation process, namely:

(1) Having training data generated under unreasonable restrictions reduces the domain of data points available for training and potentially decreases the generalization of the ML models. (2) The generated training data can result in cases where the same input leads to a different output (or target).


In this section, we first explain the choice of the benchmark and then discuss the issues around generating the training data from it. 

\subsection{Dataset Choice}

We use the JOB~\cite{JOB_leis2015good} and STACK~\cite{marcus2021bao} benchmarks for all the experiments in this paper. Sourcing data from the IMDB and StackExchange, respectively, both datasets reflect natural challenges in real-world workloads. Moreover, a recent paper~\cite{zhu2023lero} claims that the JOB benchmark is the most challenging one for LQOs, and the majority of current methods use JOB and/or STACK. We do not use the STATS-CEB benchmark suggested in \cite{han2021CEbenchmark}, as it was originally developed for challenges in cardinality estimation as opposed to end-to-end query optimization, which is the focus of this paper. We also do not use the TPC benchmark family~\cite{tpc_benchmark_doc}, as it has underlying assumptions of multivariate uniformity, which does not create reasonable challenges for LQO methods.

\subsection{Reduced Complexity of Query Plans}

During our evaluation of LQOs, we noticed that some authors suggest severely reducing the number of possible physical plans by, for example, disabling nested loop joins (as has been done in \cite{JOB_leis2015good}). This might yield improvements for some queries but solves the query optimization challenge by using a data-dependent solution at the cost of reduced generalizability. In general, we observe from some queries that limitations such as disabling specific scan or join methods, non-exact optimization, or join tree types lead to a possible increase in the chances of finding a sub-optimal plan.




E.g. the PostgresPro Community \cite{rogov2022joinmethods} discussed that any of the join methods could have an advantage over others depending on the selectivity of subqueries. The authors of \cite{marcus2021bao} show experimentally that disabling nested loop joins in PostgreSQL can improve the performance of query 16b or harm the performance of query 24b.

For bitmap and tid scans, the Genetic Query Optimizer (GEQO), and bushy trees, we provide extensive experiments producing the counter-examples in Sections \labelcref{sec:bitmap_tid_experiment,sec:geqo_experiment,sec:analysis_of_query_plans} respectively.

\subsection{Invariant Training Data Generation}
\label{sec: invariant_data_gen}

The data used as an input into LQO ML models, which all have either a reward or a prediction value, has a canonical view of $(D, y)$-pairs: $D$ refers to the vector of \emph{feature variables}, consisting of either an independent set of variables $X$ for supervised methods, or $(s, a, s^{'})$ - a set of \emph{state}, \emph{action} and \emph{next action}, respectively, for RL methods. $y$ is a \emph{target variable}, which is either the \emph{query latency, cost}, or the \emph{ranking} depending on the ML model used. 

In this subsection, we discuss why both types of variables are subject to the absence of invariance during training data generation.

\subsubsection{Feature Variables: Dynamic Optimization}

The vast majority of LQOs use the pg\_hint\_plan extension~\cite{pg_hint_plan_doc} to \emph{force PostgreSQL to execute an explicit query plan} rather than using a plan predicted by the built-in query optimizer. 

However, one should not expect that a plan with its hints is really executed. This is due to the dynamic updates of the plan during execution \cite{hellerstein2000dynamic}, referred to as \emph{dynamic optimization}. All the LQOs we evaluated force the DBMS to execute their plans during the stage of plan encoding, hence potentially training on incorrect data. 

Dynamic optimization could also be the reason for a possible discrepancy between the executed plan and the output provided by \texttt{EXPLAIN}. This means that LQOs, which rely on the cardinality estimations from \texttt{EXPLAIN}, potentially introduce significantly inaccurate estimations.

\emph{Recommendation}: This could be mitigated via a \emph{direct RL approach}, where the DBMS is treated as a "black box". The objective function is directly maximized via gradient descent without the need to learn transition probabilities (i.e., the stochastic behavior of the DBMS) and without the need to solve Bellman equations~\cite{guan2021direct}.

\subsubsection{Dependent Variables: Cold vs. Hot Cache}

If a query is executed several times, the executing time decreases due to reading pre-calculated information from previous runs (hot cache) instead of creating everything from scratch (cold cache). We want to create a situation that yields comparable and consistent results for every query.
Hence, the cache status should be either fully cold or hot, i.e., when all potential caching has been performed. No intermediate "warm" cache should be allowed. 

However, it is unreasonable to expect a full cold cache situation \cite{levandoski2013identifying}.
Moreover, it is an ethical question if it is fair enough to run queries with a cold cache, considering that it disables all the optimization techniques that the DBMS has based on cache buffers. 

\emph{Recommendation}: Taking into account potential correlations of queries inside workloads like JOB due to the use of base templates/patterns, we believe that \emph{forcing a hot cache setting is fairer}, as it mitigates the influence of previously executed queries on the execution time of any particular query. The way of achieving a hot cache setup is discussed in detail in Section \ref{sec:hot_cold_cache_experiments}; conceptually, it is a consecutive run of the same query until the latency converges.

\section{Query \& Plan Encoding}
\label{sec:query_and_plan_encoding}

In this section, we discuss which information can be extracted from SQL queries and their physical plans as input to the ML model. Moreover, we explain which principles should be followed so that LQO models are trained smoothly.

The recent LQOs, to the best of our knowledge, are all \emph{query-driven methods} in contrast to data-driven methods used for cardinality estimation~\cite{benjamin2020deepdb, yang2020neurocard}). In other words, LQOs use queries as an indispensable proxy to the data underneath the DBMS. It implies that the encoding schema for a query should be both expressive and robust. We will now discuss the \emph{principles of encoding robustness and expressiveness} and how we can achieve them.

\subsection{Encoding Robustness}

\begin{table*}[h!t]
\caption{Main encoding components of LQOs. We distinguish between \emph{query encoding} and \emph{plan encoding}. Both Bao and LOGER provide hints about what types of joins not to use. Bao also provides hints for scan types.}
\label{tab:lqos_comparison}
\resizebox{\textwidth}{!}{%
\begin{threeparttable}
\centering
\begin{tabular}{l|cccc|ccc|cccccc}
\hline
\multirow{2}{*}{LQO} & \multicolumn{4}{c|}{Query Encoding} & \multicolumn{3}{c|}{Plan Encoding} & \multicolumn{6}{c}{Training Specifics}\\
\cline{2-14}
                     & \begin{tabular}[c]{@{}c@{}}Adjacency \\ Matrix\tnote{1}\end{tabular} & \begin{tabular}[c]{@{}c@{}}Numerical \\ Attributes\tnote{2}\end{tabular} & \begin{tabular}[c]{@{}c@{}}Text \\ Attributes\end{tabular} & \begin{tabular}[c]{@{}c@{}}Encoding \\ Aggregation\tnote{*}\end{tabular} & \begin{tabular}[c]{@{}c@{}}Join \\ Type\end{tabular} & \begin{tabular}[c]{@{}c@{}}Scan \\ Type\end{tabular} & \begin{tabular}[c]{@{}c@{}}Table \\ Identifier\tnote{3}\end{tabular} 
                     & Data+\tnote{4}
                     & ML Model\tnote{*}
                     & \begin{tabular}[c]{@{}c@{}}Plan \\ Processing\end{tabular}
                     & \begin{tabular}[c]{@{}c@{}}Model \\ Output\end{tabular}
                     & Testing\tnote{*}
                     & \begin{tabular}[c]{@{}c@{}}DBMS \\ Integration\end{tabular}
                     \\ 
\hline
Neo~\cite{marcus12neo}                  & \checkmark                                                             & cardinality                                                                               & word2vec                                                   & stacking                                                                                  & \checkmark                            & \checkmark                            & \checkmark       &  -               & Regression          & Tree-CNN &              Plan &              Static &              -                                                      \\
RTOS~\cite{yu2020rtos}                 & \checkmark                                                             & filters                                                                                   & cardinality                                                & FC + pooling                                                                              & -                                                    & -                                                    & \checkmark          &  -               & Regression          & Tree-LSTM &              Plan &              CV &              -                                                   \\
Bao~\cite{marcus2021bao}                  & -                                                                                     & -                                                                                         & -                                                          & -                                                                                         & \checkmark                            & \checkmark                            & -                                   &  \checkmark       & Regression          & Tree-CNN &              Hint set &              Time Series &              \checkmark                                                   \\
Balsa~\cite{balsa_Yang2022}                & \checkmark                                                             & cardinality                                                                               & cardinality                                                & stacking                                                                                  & \checkmark                            & \checkmark                            & \checkmark                                         &  -               & Regression          & Tree-CNN &              Plan &              Static &              -                    \\
Lero~\cite{zhu2023lero}                 & -                                                                                     & -                                                                                         & -                                                          & -                                                                                         & \checkmark                            & \checkmark                            & \checkmark                         &  \checkmark       & LTR                 & Tree-CNN &              Plan &              Static &              \checkmark                                    \\
LEON~\cite{chen2023leon}                 & \checkmark                                                             & cardinality                                                                               & cardinality                                                & stacking                                                                                  & \checkmark                            & \checkmark                            & \checkmark                                    &  -               & LTR                 & Tree-CNN &              Plan &              Static &              -                         \\
LOGER~\cite{chen2023loger}                & \checkmark                                                             & filters                                                                                   & cardinality                                                & FC + pooling + GT                                                                         & \checkmark                            & -                                                    & \checkmark                                 &  -               & Regression          & Tree-LSTM &              Hint &              Static &              -                            \\
HybridQO~\cite{yu2022hybridqo}             & \checkmark                                                             & cardinality                                                                               & cardinality                                                & stacking + FC                                                                             & \checkmark                            & \checkmark                            & \checkmark                                  &  \checkmark       & Regression          & Tree-LSTM &              Plan &              Static &              -                           \\ \hline
\end{tabular}
\begin{tablenotes}
	\item[1] One-hot-encoding of the join subgraph for a particular (sub)query 
	\item[2] Filters explicitly encode >, =, and < symbols with min-max scaled filter values
	\item[3] One-hot-encoding of tables in the DBMS schema
    \item[4] Whether the method uses additional queries (outside of the provided benchmark queries) for training data generation or not
	\item[*] Stacking: assembling of several features or vectors into a single vector, Pooling: downsampling of the spatial dimensions of the input data, CV: Cross-validation on JOB, FC: Fully-connected layer in the neural network, GT: Graph transformation, LTR: Learning-to-rank, Static: Static split of JOB, Time Series: Sequential continuous testing on previously unseen queries
\end{tablenotes}
\end{threeparttable}
}
\end{table*}

Table \ref{tab:lqos_comparison} gives an overview of the main encoding components used by various LQOs. Note that we distinguish between \emph{query encoding} (information that is independent of how the query is executed) and \emph{plan encoding} (information based on the physical plan). For instance, the text attributes of the query can either be encoded based on their cardinality or by using e.g. word2vec to generate a vectorized form. Moreover, encodings can be aggregated using either stacking or pooling, sometimes with additional post-transformations. 

We notice that Bao~\cite{marcus2021bao} and Lero~\cite{zhu2023lero} do not use query encoding but only plan encoding. For instance, Bao does not identify which table is used in a particular node of the query plan, using only table cardinalities and costs. Such a representation can benefit from more schema-agnosticism and easier re-training when the database schema changes, though it violates the \emph{principle of invariance}~\cite{la1976stability}.

Let us consider the following thought experiments. Applying different filters in a query can result in the same cardinality for the same table. Similarly, tables with the same cardinalities can have the same encoding. In an ideal setting, we would want a unique 1-to-1 mapping between the feature variables $D$ and the latency or cost $y$ of a query and its given plan. However, the query latency is volatile and differs between multiple executions so that the plan encoding will instead result in a 1-to-many mapping of ($D$, $y$)-pairs.


Moreover, even having the query encoding as an additional input cannot guarantee invariance under a single cardinality encoding of the attributes. As we have discussed in the example above, applying different filters for a given column can result in the same cardinality estimation, i.e., leads to the loss of invariance.


\emph{Recommendation}: To avoid spoiling the training process by not having the 1-to-1 mappings for ($D, y$)-data pairs, one can \emph{use the embeddings instead of single value representations}, e.g., embeddings for text attributes like in Neo (see Figures 12 and 13 in the original paper~\cite{marcus12neo}), and explicit vectorization of filters like in RTOS~\cite{yu2020rtos}.

\subsection{Encoding Expressiveness}

The final set of features should clearly reflect both the global and local context. In query optimization, the global context is the \emph{query} (as it does not change throughout the physical plan space search), and the local context is the \emph{query plan}. This concept comes from Graph CNNs~\cite{kipf2016gcnns}. The basic idea is that applying more rounds of convolutions in the neural architectures will result in a graph node embedding with more global graph context and less local context. 

Continuing the idea of using graph NNs, graph transformers~\cite{dwivedi2021gt} are used in LOGER~\cite{chen2023loger} in an adjacent context for query encoding aggregation. On the other hand, methods like Bao~\cite{marcus2021bao} and Lero~\cite{zhu2023lero} are missing the query encoding part, which increases the probability of converging to a local optimum~\cite{guyon2006features}.

\emph{Recommendation}: We would suggest \emph{using both the query and the plan encoding}, which will result in better convergence.


\section{Training Learned Query Optimizers}
\label{sec:lqo_training}

In this section, we discuss how the "brains" of LQOs work and what conditions should be met to make them work as expected. The key feature of recent LQOs is the possibility to learn the entire query optimizer process with the help of ML models. From Table \ref{tab:lqos_comparison}, it is visible how different the training pipelines are among LQOs. For example, a query plan having a tree representation structure implies two possibilities when processing: some can treat it as an image and apply Tree Convolutions~\cite{mou2016treeconv}, others treat it as a sequence of node pairs (i.e., text) and apply a Tree-LSTM~\cite{tai2015treelstm}. However, there is still no common ground, e.g., for the performance analysis during model training or the choice of the training method. In this section, we discuss the most widespread issues of LQOs at the training phase.

\subsection{Avoiding ML Model Overfitting}

Overfitting is a typical ML problem when the model performance improves on the training data and at the same time deteriorates on the validation data \cite{webb2010ml}. From the definition, it is clear that RL-based methods do not suffer from this problem because they learn an optimal policy by maximizing or minimizing a non-stationary objective function that depends on the action policy itself. However, RL methods might get stuck in a sub-optimal policy without enough exploration~\cite{sutton1998rl}. Contrarily, classical supervised methods are prone to converge to a suboptimal solution.

To avoid overfitting, commonly \emph{hyperparameter tuning via cross-validation (CV)}, \emph{early stopping} and \emph{regularization} are applied. Regularizations like, e.g., dropout~\cite{wan2013dropout} are straightforward and simply increase the number of hyperparameters that need to be tuned, though other techniques are harder to tweak. Among recent LQOs, only RTOS  applies CV to measure final aggregated performance metrics, though this does not help choose the final model. Balsa uses early stopping with performance improvement on the non-fixed validation set. LEON is doing a similar early stopping procedure, though using accuracy as a target metric. Bao uses a continuous "time series" testing of the model on previously unseen queries.

\emph{Recommendation}: For RL methods, one can still \emph{use hyperparameter tuning} as it would also help improve the general model performance. For QOPs, accuracy for both cost and latency is a suboptimal quality metric as we do not know the optimal plan in advance (at least for higher-order joins). Thus, using accuracy as an early-stopping or cross-validation criterion is undesirable. The holdout data should be fixed (not CV, not "time series"), as the measurement on it should be comparable~\cite{goodfellow2016dl}. 

\subsection{Changing Target Variables On-the-Fly}

Query optimization has interesting specifics regarding the target to be optimized, which could either be a cost or a latency. It results in finding a \emph{trade-off between speed} (as costs could be quickly estimated by an arbitrary cost model) \emph{and accuracy} (as latency gives the exact value for how long the query takes to execute). Some methods like HybridQO take advantage of both by first training the model that suggests plans based on cost and then training another model that chooses between candidates based on latencies. At the same time, methods like RTOS, Balsa, Lero, and LEON try to use a single predictive model that first pre-trains using costs and then continues training with latencies. A key issue of this approach is that latencies and costs have significantly different numerical properties, and any progress made in the pre-training phase is lost, as the model needs to adapt to an entirely new scale and variance (i.e. deviation from the mean) of the target values~\cite{brownlee2020data}.


\emph{Recommendation}: You can \emph{exchange the cost and latency on-the-fly during the training when using learning-to-rank models}, since real values are transformed into relative rankings forming the target variable~\cite{liu2009ltr}. Another approach is to \emph{use an architecture that chains the ML models} like in HybridQO, where different target variables are served to different models in the ML pipeline.

\section{Evaluating Learned Query Optimizers}
\label{sec:lqo_evaluation}

In this section, we outline the importance of choosing the right test set, how this decision influences the model's measured performance, and the concept of covariate shift.

\subsection{Test Set Choice}

The \emph{train/test split} is a cornerstone of any supervised method. This split is used to differentiate between which part of the data an ML model is allowed to see during training and which part is used to test its ability to perform on previously unseen data, measuring the generalization ability of the model. 

The extended JOB workload introduced by Neo \cite{marcus12neo} was a first attempt to test the ability of models to deal with previously unseen queries that are distinct from the original JOB queries. The queries added in Ext-JOB exhibit additional operators that are not present in JOB (such as GROUP BY or ORDER BY). Due to the nature of merge joins~\cite{kacimi2009systemr}, LQOs that prefer this join method tend to gain an advantage from including ORDER BY operators. As a result, the comparison between different methods is unfairly skewed.

\begin{figure}[h!]
    \centering
    \includegraphics[width=\linewidth, keepaspectratio]{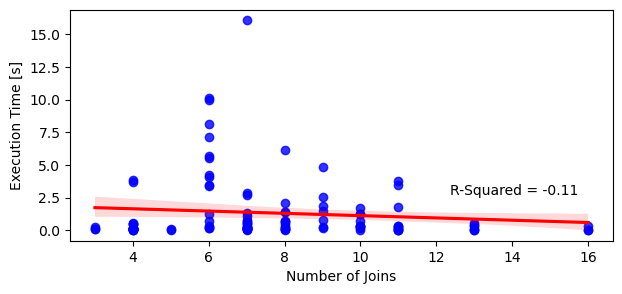}
    \caption{Scatter plot of the execution time per number of joins for all queries in JOB.}
    \label{fig:n_joins_vs_exec_time}
\end{figure}

Balsa introduced JOB-Slow, where the 19 slowest queries shape the test set, and all other queries are the training set. This intuitively simple-to-understand train/test split focuses on the queries that have the most impact on the overall execution time for a full workload. However, all the 19 queries of the JOB-Slow test set have 11 or fewer joins, while 11 queries have just 6 or fewer joins. Figure \ref{fig:n_joins_vs_exec_time} shows a scatter plot of the execution time vs. the number of joins. We observe that queries having between 6 and 11 joins have the largest execution times and thus, the highest potential for being optimized. At the same time, this is the range where non-exhaustive optimizers are typically disabled (e.g., PostgreSQL's GEQO is by default only enabled for 12 or more tables). Hence, exhaustive methods can still fully explore the space of possible plans.

Another approach for splitting queries was introduced by \cite{zhang2023adaptive}, where the authors built train/test splits based on the number of joins. For example, all queries with 3 or 4 joins form the test set, and all others form the training set. From Figure \ref{fig:n_joins_vs_exec_time} it is clear that the number of joins is an irrelevant proxy for execution time, according to a regression analysis with $R^{2}=-0.11$. Thus, splitting queries as such forms groups that are not aligned with the true optimization target, i.e., the execution time.

\emph{Recommendation:} We propose several \emph{edge cases for train/test splits to cover different areas of generalization}, namely the generalization gap and sampling out-of-distribution (see Section \ref{sec:dataset_split}).


\subsection{Covariate Shift}
Another relevant topic for evaluating LQOs is the concept of \emph{covariate shift}, i.e. a change in the database content away from how a method was trained. DBMSes tackle this challenge by continuously updating the internal statistics. For a LQO, however, a change in the database content affects how a query is encoded and thus its prediction. For example, a query about movies with a release date greater than 2022 will continuously increase its result set size, as newly released movies are added to the DBMS. 

While this topic is often mentioned in aspirational future work, methods like Bao have started to think about designing their encoding to be able to deal with covariate shift by omitting tables and column identifiers in their encoding (see Section \ref{sec:query_and_plan_encoding} for more details). However, as we show in an experiment in Section \ref{ablation_study__covariate_shift}, updated cardinality estimates in the encoding are insufficient to keep up with changing database content.

\emph{Recommendation:} We propose that future methods should \emph{include a simple experiment to measure the ability to deal with covariate shift}, as we have performed in Section \ref{ablation_study__covariate_shift}.


\section{Framework for Benchmarking Learned Query Optimizers}
\label{sec:framework}

In this section, we introduce our benchmarking framework, aiming for a comprehensive evaluation of LQOs. The objective is to conduct benchmarking in a holistic manner, ensuring a fair comparison of methods in an end-to-end setting.

To do this, our benchmark assumes a reproducible setup, particularly regarding engineering, including but not limited to (a) the content of the database underlying a benchmark workload, (b) the full code base of the LQO, (c) the version of the programming language, such as Python, and all used libraries, (d) a detailed configuration of the DBMS (unless all parameters are left on default), as well as (e) all queries and their assignment into train/test splits.

\begin{table*}[h!t]
    \caption{Overview of different PostgreSQL configurations (database tuning parameters) used in various papers of LQOs. \emph{Deviations from PostgreSQL's default values are marked} in the respective columns. Note, that the values for Neo~\cite{marcus12neo} and HybridQO~\cite{yu2022hybridqo} are missing from the table, as their configuration parameters are not publicly available.}
    \label{tab:database_configuration_table}
    \resizebox{0.9\textwidth}{!}{%
    \begin{threeparttable}
        \begin{tabular}{lccccccc}
        \multicolumn{1}{l|}{\textbf{PostgreSQL Config Parameter}} & \multicolumn{1}{c|}{\textbf{Default Values}} & \multicolumn{1}{c|}{JOB \cite{JOB_leis2015good}} & \multicolumn{1}{c|}{Bao \cite{marcus2021bao}} & \multicolumn{1}{c|}{Balsa \cite{balsa_Yang2022}, LEON \hspace{1mm}\cite{chen2023leon}} & \multicolumn{1}{c|}{LOGER \cite{chen2023loger}} & \multicolumn{1}{c|}{Lero\hspace{1mm} \cite{zhu2023lero}} & \textbf{Our Framework}         \\ \hline \hline
        \multicolumn{1}{l|}{Amount of RAM used by authors}        & \multicolumn{1}{c|}{}                        & \multicolumn{1}{c|}{64 GB}                       & \multicolumn{1}{c|}{15 GB}                    & \multicolumn{1}{c|}{64 GB}                                                                      & \multicolumn{1}{c|}{256 GB}                     & \multicolumn{1}{c|}{512 GB}                                       & 64 GB        \\ \\[-2.2ex]
        \textbf{Join Order}                                       &                                              &                                                  &                                               &                                                                                                 &                                                 &                                                                   &              \\ \hline
        \multicolumn{1}{l|}{geqo\_threshold}                      & \multicolumn{1}{c|}{12}                      & \multicolumn{1}{c|}{18}                          & \multicolumn{1}{c|}{}                         & \multicolumn{1}{c|}{}                                                                           & \multicolumn{1}{c|}{2 or 1,024}        & \multicolumn{1}{c|}{}                                             &              \\
        \multicolumn{1}{l|}{geqo}                                 & \multicolumn{1}{c|}{on}                      & \multicolumn{1}{c|}{}                            & \multicolumn{1}{c|}{}                         & \multicolumn{1}{c|}{off}                                                                        & \multicolumn{1}{c|}{off}               & \multicolumn{1}{c|}{off}                                          & off\tnote{1} \\ \\[-2.2ex]
        \textbf{Working Memory}                                   &                                              &                                                  &                                               &                                                                                                 &                                                 &                                                                   &              \\ \hline
        \multicolumn{1}{l|}{work\_mem}                            & \multicolumn{1}{c|}{4 MB}                    & \multicolumn{1}{c|}{2 GB}                        & \multicolumn{1}{c|}{}                         & \multicolumn{1}{c|}{4 GB}                                                                       & \multicolumn{1}{c|}{}                           & \multicolumn{1}{c|}{}                                             & 4 GB         \\
        \multicolumn{1}{l|}{shared\_buffers}                      & \multicolumn{1}{c|}{128 MB}                  & \multicolumn{1}{c|}{4 GB}                        & \multicolumn{1}{c|}{4 GB}                     & \multicolumn{1}{c|}{32 GB}                                                                      & \multicolumn{1}{c|}{64 GB}                      & \multicolumn{1}{c|}{}                                             & 32 GB        \\
        \multicolumn{1}{l|}{temp\_buffers}                        & \multicolumn{1}{c|}{8 MB}                    & \multicolumn{1}{c|}{}                            & \multicolumn{1}{c|}{}                         & \multicolumn{1}{c|}{32 GB}                                                                      & \multicolumn{1}{c|}{}                           & \multicolumn{1}{c|}{}                                             & 32 GB        \\
        \multicolumn{1}{l|}{effective\_cache\_size}               & \multicolumn{1}{c|}{4 GB}                    & \multicolumn{1}{c|}{32 GB}                       & \multicolumn{1}{c|}{}                         & \multicolumn{1}{c|}{}                                                                           & \multicolumn{1}{c|}{}                           & \multicolumn{1}{c|}{}                                             & 32 GB        \\ \\[-2.2ex]
        \textbf{Parallelization}                                  &                                              &                                                  &                                               &                                                                                                 &                                                 &                                                                   &              \\ \hline
        \multicolumn{1}{l|}{max\_parallel\_workers}               & \multicolumn{1}{c|}{8}                       & \multicolumn{1}{c|}{}                            & \multicolumn{1}{c|}{}                         & \multicolumn{1}{c|}{}                                                                           & \multicolumn{1}{c|}{1}                          & \multicolumn{1}{c|}{0}                                            &              \\
        \multicolumn{1}{l|}{max\_parallel\_workers\_per\_gather}  & \multicolumn{1}{c|}{8}                       & \multicolumn{1}{c|}{}                            & \multicolumn{1}{c|}{}                         & \multicolumn{1}{c|}{}                                                                           & \multicolumn{1}{c|}{1}                          & \multicolumn{1}{c|}{0}                                            &              \\
        \multicolumn{1}{l|}{max\_worker\_processes}               & \multicolumn{1}{c|}{2}                       & \multicolumn{1}{c|}{}                            & \multicolumn{1}{c|}{}                         & \multicolumn{1}{c|}{8}                                                                          & \multicolumn{1}{c|}{}                           & \multicolumn{1}{c|}{}                                             & 8            \\ \\[-2.2ex]
        \textbf{Scan Types}                                       &                                              &                                                  &                                               &                                                                                                 &                                                 &                                                                   &              \\ \hline
        \multicolumn{1}{l|}{enable\_bitmapscan}                   & \multicolumn{1}{c|}{on}                      & \multicolumn{1}{c|}{}                            & \multicolumn{1}{c|}{}                         & \multicolumn{1}{c|}{off}                                                                        & \multicolumn{1}{c|}{}                           & \multicolumn{1}{c|}{}                                             &              \\
        \multicolumn{1}{l|}{enable\_tidscan}                      & \multicolumn{1}{c|}{on}                      & \multicolumn{1}{c|}{}                            & \multicolumn{1}{c|}{}                         & \multicolumn{1}{c|}{off}                                                                        & \multicolumn{1}{c|}{}                           & \multicolumn{1}{c|}{}                                             &             
        \end{tabular}
        \begin{tablenotes}
            \item[1] GEQO is only turned on for Bao and when PostgreSQL fully controls the query execution. 

        \end{tablenotes}
    \end{threeparttable}
    }
\end{table*}

\subsection{DBMS Configuration \& Database Tuning}
\label{subsec:dbms_config}

For analyzing query execution times, both the used hardware and the DBMS configuration greatly impact the comparability of LQOs. We will now analyze the major parameter settings systematically.

Table \ref{tab:database_configuration_table} gives an overview of the different parameter settings used in various publications, compared to the default values of PostgreSQL, as well as the suggested setting for the Join Order Benchmark \cite{JOB_leis2015good}. Note that the configurations for Neo~\cite{marcus12neo} and HybridQO~\cite{yu2022hybridqo} are omitted from Table \ref{tab:database_configuration_table}, as their code (Neo) and database configuration (HybridQO) are not publicly available.
A further observation is that only Balsa and LEON published the full DBMS configuration file among their artifacts.
We have categorized the parameters into the following groups:

\emph{Join Order:} The join order is typically forced through libraries such as \texttt{pg\_hint\_plan}~\cite{pg_hint_plan_doc}, though PostgreSQL can also be made to follow the explicit order given in the SQL statement by setting \texttt{join\_collapse\_limit} to 1.
The genetic query optimization algorithm (GEQO) of PostgreSQL is used for queries with large number of joins, by default 12 or more. It can either be disabled by setting \texttt{geqo\_threshold} to a value larger than the number of joins in a workload or disabled completely with the \texttt{geqo} parameter.

\emph{Working Memory:} The default values for PostgreSQL's memory are small. Given the amount of RAM available today, increasing the working memory and buffer sizes is advisable.
Balsa drastically increases the working memory (\texttt{work\_mem}) from 4 MB to 4 GB, while Bao and Neo keep the default value, despite the proposed 2 GB by \cite{JOB_leis2015good}. Similarly, for the \texttt{shared\_buffers}, Balsa uses a much larger buffer at 32 GB compared to the 4 GB recommendation that Bao and Neo use. LOGER further increases the \texttt{shared\_buffers} value to 64 GB, though their machine also has more RAM available.

Note that the amount of \texttt{work\_mem} is available to all workers in parallel query execution, that means for $N$ amount of workers, the \texttt{shared\_buffers} should be at least $N \times$ \texttt{work\_mem}. Furthermore, all methods use the default cache size (\texttt{effective\_cache\_size}) of 4 GB, ignoring the recommendation to increase it to 32 GB by \cite{JOB_leis2015good}. Increasing its value in our configuration from 4 to 32 GB reduced the planning time for a handful of outlier queries significantly (from up to 3 seconds to below 100 milliseconds).





\emph{Parallelization:} These parameters define the number of workers and processes used during query execution. To fully utilize a multi-core system, Balsa increases the number of worker processes \texttt{max\_worker\_ processes} to match \texttt{max\_parallel\_workers}. While increasing the number of parallel workers can speed up query execution, the amount of required compute resources also increases significantly. LOGER and Lero take a different approach, disabling any parallel query execution completely.

\emph{Scan Types:} These parameters directly change the types of scans that are being used by PostgreSQL and significantly alter the toolset available for query execution. Only Balsa and LEON change these values by disabling both bitmap and tid scans, while neither paper offers an explanation for taking this approach. 


\subsection{Dataset Split}
\label{sec:dataset_split}

\begin{figure*}[h!t]
  \centering
  \includegraphics[width=\textwidth, keepaspectratio]{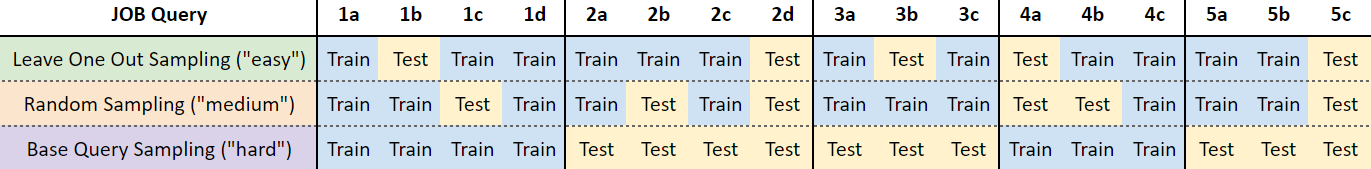}
  \caption{Overview of different dataset split sampling types for JOB: Leave One Out Sampling (top), Random Sampling (middle), and Base Query Sampling (bottom).
  For instance, Base Query 1 has 4 variations: \texttt{1a}, \texttt{1b}, \texttt{1c} and \texttt{1d}. 
  }
  \label{fig:split_type_overview}
\end{figure*}

The way the dataset is split into training and test sets has a significant impact on the performance of a trained model. While it is advisable that both sets contain data from a similar distribution, we have to be careful to avoid leaking information from one set to the other. More specifically, the Join Order Benchmark queries are deduced from 33 different base queries (or templates), and the full 113 queries are made up of between 2 and 6 variations of each base query (denoted as \texttt{1a}, \texttt{1b}, \texttt{1c}, ...). Variants of the same base query share the same tables and joins but differ in filter statements. These differences can be different filter values (e.g. \texttt{production\_year < 2000} vs. \texttt{production\_year = 2023}) or applying filters on other columns (e.g. \texttt{genre = 'horror'} vs. \texttt{name LIKE \%an\%}). The queries in the STACK~\cite{marcus2021bao} dataset also follow the same pattern of 16 base queries across 6,191 queries (with 100 to 1,010 variations per query).

Generating queries from templates introduces a strong correlation in the structure of the optimal join plan for some, but not all, queries. To measure the effect of potential data leakage, we propose the following sampling techniques to generate dataset splits (see Figure \ref{fig:split_type_overview} for a visual example of training and test set assignments):

\noindent
(1) \textbf{Leave One Out Sampling} extracts exactly one variant of each base query into the test set. All other variants of the base query are contained in the training set. This split maximizes the amount of information that can potentially be leveraged from the training onto the test set. We expect this split to be the \emph{easiest to learn}.

\noindent
(2) \textbf{Random Sampling} distributes all queries randomly into train and test sets, ignoring any base query or template affiliations. This is a \emph{medium difficulty} sampling, and it can be applied to any workload, as there is no requirement for the existence of \textit{base query families}.

\noindent
(3) \textbf{Base Query Sampling} keeps all queries of the same base query either in the training or the test set. This ensures that the intra-family similarity of the query structure does not leak from the training set into the test set. We believe this to be the \emph{most difficult} split, as a model cannot apply the join structure learned from one variant of the same base query to another.


    

\subsection{Measuring Query Executions}
\label{sec:hot_cold_cache_experiments}

As LQOs are all evaluated by the runtime of queries in a workload, and some LQOs directly predict the execution time for a given physical plan, it is vital that runtime measurements are as consistent as possible. One of the primary reasons for high variance in executing the same query is caused by the buffer and cache states in the DBMS. For example, when the same query is executed twice one after another, the first run generally takes longer than the second one. As buffers and caches switch from cold to hot cache, runtimes become more consistent. In the ideal scenario, we could execute every query many times to achieve a robust measurement. However, every additional execution after the first one takes additional time that is not spent on executing other queries, costing valuable compute resources. We experimentally determined that executing queries \textit{3 times} and taking the third execution gives the most stable results without incurring an unnecessary amount of execution overhead (see Section \ref{sec:analysis_of_successive_query_execution_times} for more details on the experiment).

\section{Experiments}
\label{sec:experiments}

In this section, we present our extensive evaluation of LQOs on the Join Order Benchmark (JOB) and STACK. First, we give an overview of the setup and hardware used; then, we discuss different approaches to generate train/test splits. Finally, we show the results of our experiments with a number of ablation studies.

\subsection{General Setup}

\subsubsection{Software and Hardware}

All our experiments were conducted using PostgreSQL version 12.5 by measuring the query execution time through \texttt{EXPLAIN} \texttt{ANALYZE} calls, using both execution and planning time. In addition, we also include the inference time for LQOs. Measurements are taken by executing the same query three times and taking the last query execution (hot cache).

Our instance of PostgreSQL is configured largely with default parameters in mind, closely following the configuration used by Balsa \cite{balsa_Yang2022}. In comparison, we reenabled both bitmap and tid scans and increased the \texttt{effective\_cache\_size} from 4 to 32 GB. The main differences to PostgreSQL's default can be seen in Table \ref{tab:database_configuration_table} and primarily include changes to the memory configuration and an increased amount of parallel workers. In addition, we disabled the \texttt{AUTOVACUUM} feature, as the query workload is stable and \texttt{ANALYZE} is run once after loading all data into PostgreSQL, taking 3 minutes for IMDB (JOB) and 16.5 minutes for STACK.

We have decided to follow the configuration of the Balsa experiments, as they include memory settings that strongly follow the best practices guide proposed by PostgreSQL \cite{pg_best_practices} and the suggestions of Leis et al. \cite{JOB_leis2015good}. Furthermore, Balsa is the first method to increase the number of available workers processes from 2 to 8, given the typical machines with many CPU cores. We further change the \texttt{effective\_cache\_size} parameter in line with the best practices of PostgreSQL and re-enable both bitmap and tid scans.

For the Join Order Benchmark, the authors of Balsa added two additional indexes on the \texttt{subject\_id} and \texttt{status\_id} columns of the \texttt{complete\_cast} table, compared to the indexes provided by \cite{JOB_leis2015good}. We also include the additional indexes in our experiments. The experiments were run inside Docker containers, using a Tesla T4 GPU, 64 GB of RAM, and 16 CPU cores.

\subsubsection{Query Workload}

We evaluate various LQOs on the JOB and STACK workloads. Both workloads are highly relevant in recent literature and have been used in most of the evaluated LQO methods.
For JOB \cite{JOB_leis2015good}, we use the 113 queries provided. The STACK \cite{marcus2021bao} workload includes 6,191 queries across 16 base queries, which we down-sampled to 14 base queries\footnote{Templates 9 and 10 are removed in accordance with Balsa~\cite{balsa_Yang2022}, where the authors report a limitation in the pg\_hint\_plan extension in dealing with views and subqueries.} with 8 randomly sampled variations, each. This allows the methods to be trained and evaluated using a similar amount of data for JOB and STACK, leaving the models at a similar level of statistical power. 

\subsubsection{Dataset Split}

For our experiments, we generated the train/test splits by uniformly sampling across all queries (Random Sampling), the base queries (Base Query Sampling), or the variants of each base query (Leave One Out Sampling). For the Random splits and Base Query splits, we used an 80-20 ratio between training and test sets.
The dataset splits are sampled once and shared across all the evaluated methods. Detailed listings of the training and test sets for all splits can be found in our code repository\footnote{See our code here: \url{\vldbavailabilityurl}}, along with the hyperparameters of all methods.

\subsubsection{Additional Noteworthy Changes}

As we evaluate the LQO methods under our unified framework, there are differences to the experiments conducted by the authors of the methods (see previous sections). Hence, direct comparisons to prior results are impossible. 

In addition, Bao was originally trained on 2,500 newly generated queries in the JOB workload style. In our experiments, Bao was only trained on the training set of the respective train/test splits and has seen the training queries multiple times. For LEON, we have limited the amount of real time spent on training to twice the time it took Balsa to finish training, i.e., 120 hours. This time budget likely reduces the performance of LEON, but as shown in Section \ref{sec:training_time}, the inference time heavily dominates its overall runtime, not just the execution time.

\begin{figure*}[h!t]
  \centering
  \includegraphics[width=\textwidth, keepaspectratio]{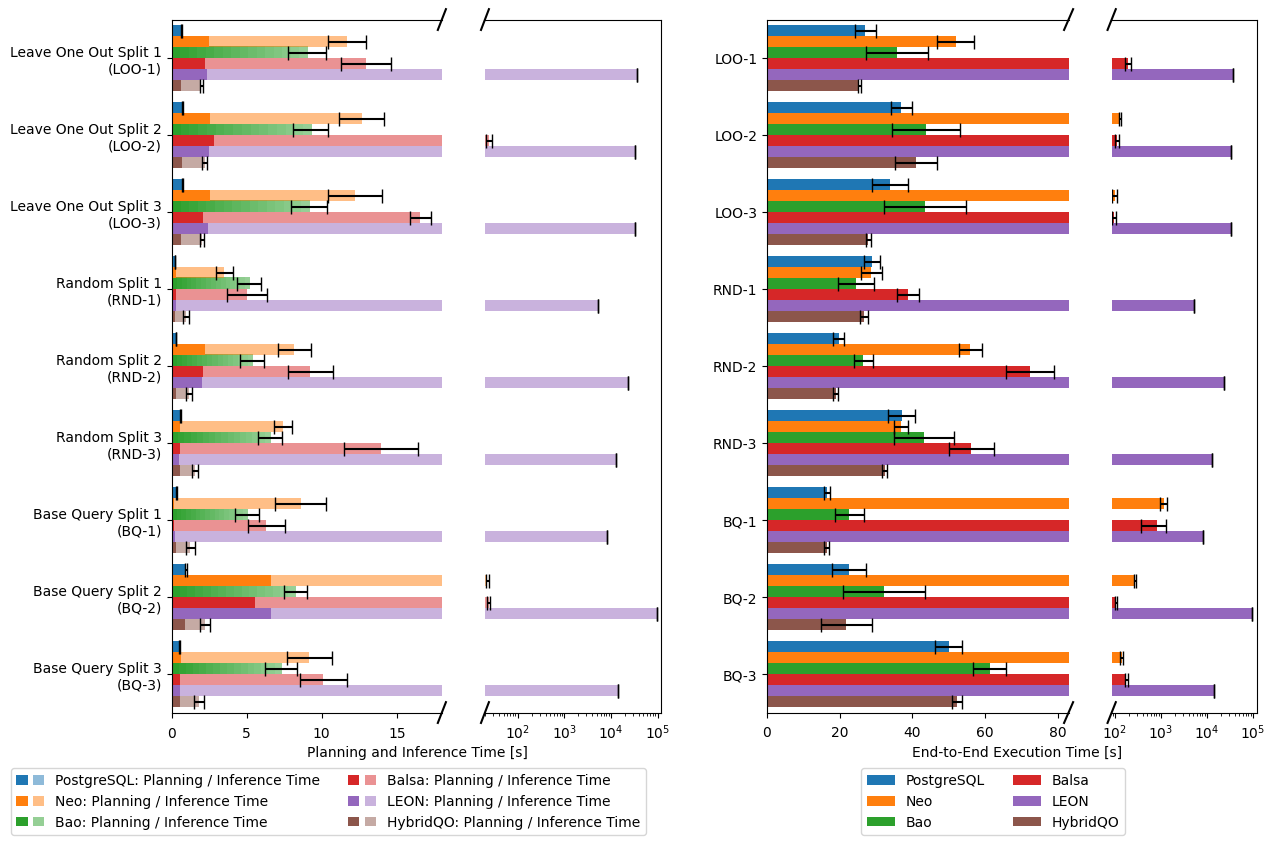}
  \caption{Comparative overview of each method's performance on the \textit{test set} of various dataset splits on the Join Order Benchmark (JOB). The figure on the left depicts the planning time (darker colour) and inference time (lighter colour), respectively. Note that Bao runs inside PostgreSQL as an extension, and its inference time is directly added to the planning time. The figure on the right side shows the execution times on the same train/test splits. Please observe that the x-axis of both figures is divided.}
  \label{fig:results_detailed__test__JOB}
\end{figure*}

\begin{figure*}[h!t]
  \centering
  \includegraphics[width=\textwidth, keepaspectratio]{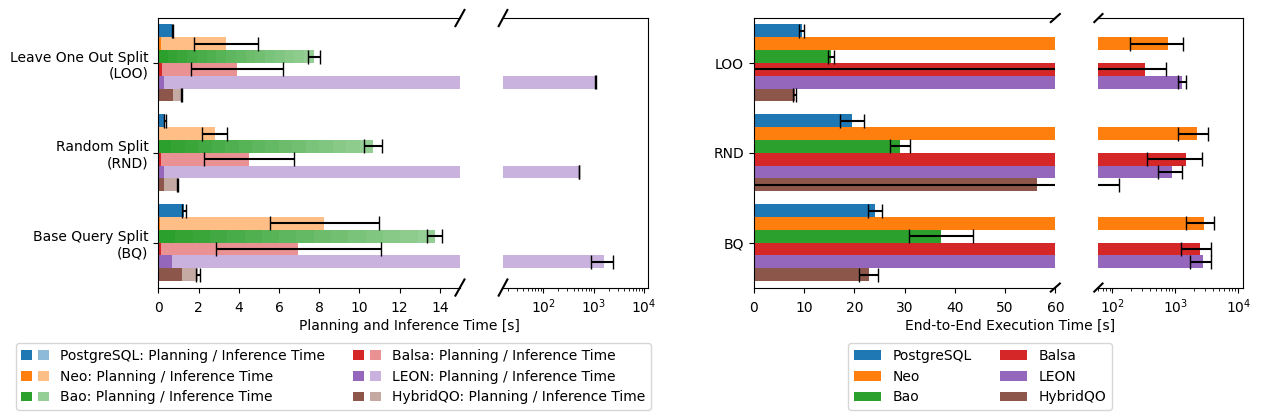}
  \caption{Comparative overview of each method's performance on the \textit{test set} of various dataset splits on STACK.}
  \label{fig:results_detailed__test__STACK}
\end{figure*}

\subsection{Comparison of Current State-of-the-Art Learned Query Optimizers}
\label{subsec:comparison_sota}

In this section, we analyze the performance of current state-of-the-art methods for LQOs (namely Neo\footnote{Since the code of Neo is not publicly available, we have used the re-implementation in Balsa. The original authors of Neo have checked it and confirmed to us that it is suitable for benchmarking.}~\cite{marcus12neo}, Bao~\cite{marcus2021bao}, Balsa~\cite{balsa_Yang2022}, LEON~\cite{chen2023leon} and HybridQO~\cite{yu2022hybridqo}) compared to PostgreSQL as our baseline.
We do not include RTOS~\cite{yu2020rtos}, Lero~\cite{zhu2023lero} and LOGER~\cite{chen2023loger} in our experiments because they are either (a) unavailable, (b) require to disable parallelization in query executions because multiple queries are run in parallel or, (c) require to invest an extensive amount of engineering to enable these methods to parse the EXPLAIN output.


\subsubsection{End-to-End Performance}
\label{sec:end_to_end_Performance}

For all algorithms, we report a variety of time measurements defined as follows:

\noindent
(1) \textbf{Inference Time:} This measure includes all time that an LQO spends to encode a query, iterate over variations of query plans, gather cost information to guide further decisions, and finally, use an ML model to generate predictions. After the inference time has passed, a given SQL query is ready to be sent to PostgreSQL with hints on which scan or join types to use and in which order.

\noindent
(2) \textbf{Planning Time:} Once PostgreSQL receives a query, it spends an amount of time on planning the query before a final physical plan is generated and sent for execution. For LQOs with an extension running inside PostgreSQL, typically, the inference time is reported as part of the planning time.

\noindent
(3) \textbf{Execution Time:} Encompasses the amount of time spent by PostgreSQL to execute the query and gather the result set. 

\noindent
(4) \textbf{End-to-end Execution Time:} A combination of the previous three-time measurements, measuring how long a method takes to devise a query plan to execute and how much time PostgreSQL spends to get the result from the database. We believe this measurement to be the primary objective for optimization.

    
    
    

We do not include the network latency in our inference, planning and execution times, as LQOs cannot influence it directly. While the network latency can be a significant amount of time (notably for fast queries), optimizing for it is beyond the scope of this evaluation.

We increase the difficulty iteratively across experiments, starting with the leave one out sampling, then the random sampling and finally, the base query sampling that generated the train/test splits. All queries were executed three times. The planning and execution times have been taken from the third execution. 

Figures \ref{fig:results_detailed__test__JOB} and \ref{fig:results_detailed__test__STACK} present the performance on JOB and STACK across all three sampling methods and their individual train/test splits. In summary, \textbf{PostgreSQL generally performs best, followed by HybridQO, then Bao, Neo, Balsa, and finally LEON}.
However, PostgreSQL fails to eclipse all methods on all splits by a statistically significant margin.
In particular, HybridQO and Bao achieve comparable results on most train/test splits, with HybridQO outperforming PostgreSQL on the leave one out split of STACK.

Let us first take a look at the results on JOB. For the \textit{leave one out sampling}, which we consider to be the easiest train/test split, PostgreSQL, Bao and HybridQO execute the test queries in around 30 seconds. However, Bao spends 8.5 seconds longer to plan queries, resulting in a 25\% slower end-to-end execution time. Bao's larger confidence interval gives a first hint that it has found plans that are generally faster than PostgreSQL, but they are not speeding up the execution time enough to have an advantage. HybridQO, on the other hand, finds plans that are 2.5 seconds faster while only spending 1.4 seconds for inference, allowing it to outperform PostgreSQL significantly on the third split.

LEON is the fourth fastest method by execution time at 58 seconds, but its inference time is around 9.6 hours long, making its use impractical for interactive querying (with more complex queries requiring proportionally more inference time to complete). 

The overall fourth fastest method is Neo at 93 seconds, followed by Balsa at 134 seconds with 286\% and 411\% slower end-to-end execution times compared to PostgreSQL, respectively.

For the \textit{random sampling}, i.e., the medium difficulty train/test split, PostgreSQL and Bao remain competitive with each other, with Bao achieving even a lower execution time of 25 vs. PostgreSQL's 28 seconds. However, this is not a statistically significant difference. Including the inference and planning times as well, Bao is again at a slight disadvantage. Similar to the previous split, HybridQO matches PostgreSQL and even significantly outperforms it and Bao on the third split with 24 seconds. Compared to the leave one out sampling, both Neo and Balsa achieve 2-3$\times$ faster end-to-end execution times, reaching comparable results for PostgreSQL on 2 out of the 3 train/test splits using this sampling. LEON struggles with these queries and two queries timeout (26b and 32b) in two separate splits, leading to a drastic increase in the execution time. However, given the large inference time of on average of 3.8 hours, this has little impact on its overall ranking.

Finally, let us examine the \textit{base query sampling}, i.e., the most difficult sampling technique. For the first time, Bao only achieves comparable results to PostgreSQL on 1, and HybridQO on all, out of the 3 train/test splits, confirming the increased difficulty. Neo and Balsa struggle particularly with base query split 1. 3 queries of the test set timeout for Neo, and 15 queries across both train and test set for Balsa. LEON, however, can largely match PostgreSQL's direct execution time if the method can overcome its inference time.

The results on the STACK dataset (see Figure \ref{fig:results_detailed__test__STACK}) largely confirm the results of JOB. A major difference is the inference time of LEON, which is one order of magnitude lower given the generally lower number of joins in the STACK queries compared to JOB. Neo, Balsa and LEON all suffer greatly from significant amounts of timed out queries. Unlike in JOB, Bao is unable to match the performance of PostgreSQL on STACK due to more complicated SQL features in the STACK queries, leading to much longer inference times. HybridQO outperforms PostgreSQL on the leave one out split, is comparable on the base query split but also suffers from timed out queries on the random split, hinting at problems of robustness.

In summary, we see the \emph{significant impact of the inference time on the overall end-to-end execution time}. While there are methods that, on some train/test splits, perform comparable to PostgreSQL or even slightly outperform it, these results show the \emph{importance of how queries are split for training}. Furthermore, it is vital that evaluations include the inference time, as it strongly shapes the ranking between methods compared to the execution time alone.

\begin{figure}[h!]
  \centering
  \begin{minipage}[b]{\linewidth}
    \centering
    \includegraphics[width=\linewidth, keepaspectratio]{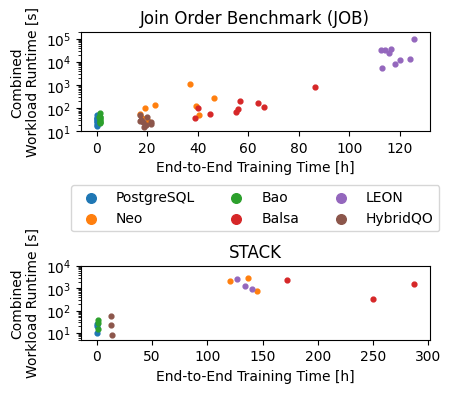}
  \end{minipage}%

  \caption{Comparison of the end-to-end training time against the combined workload runtime for both workloads, where each dot represents a model from a different split.}
  \label{fig:training_time_comparison}
\end{figure}

\subsubsection{Training Time}
\label{sec:training_time}

After comparing the query performances, we also take a look at the amount of time to train a model. While the definition of training time is sometimes unclear, we intend to take a holistic look with an \textit{end-to-end training time}, that is, including (a) the time spent collecting query results from the DBMS, (b) any time spent training the model, (c) the ongoing evaluation of the current model's performance, and (d) any pre- or postprocessing, initialization, and artifact generation. In short, the \emph{full amount of time spent from starting the training procedure until it terminates}.

We make this distinction not to penalize additional logging or more frequent checks on the model performance but to get a fairer overall comparison. For example, a method might have a very quick training period but spend a lot of time querying training data from the DBMS, while another method needs fewer database queries but uses a more complex NN architecture that spends more time in weight updates. The overall amount of time spent also informs how often a model could be retrained given a time budget.

In Figure~\ref{fig:training_time_comparison} we compare the end-to-end training times on the x-axis and the combined workload runtimes (the sum of the end-to-end execution times of all queries in the workload) on the y-axis. Each dot represents one train/test split. For example, one orange dot could be the Neo method on random split 2.

Since PostgreSQL's optimizer does not require any inherent training acting as a baseline, its end-to-end training time is set to zero. Among the evaluated methods, Bao requires the least amount of time on JOB at around 2 hours, HybridQO around 20 hours, Neo between 20 and 40 hours, Balsa between 40 and 85 hours, and LEON from 110 to 130 hours. For STACK, Bao trains for 2 hours, HybridQO for 12 to 14 hours, Neo and LEON between 120 and 140 hour, and Balsa between 170 and 290 hours.

As a naive assumption, one would expect to see a performance increase as more time is spent during training,
but we observe exactly the opposite behaviour: Methods that have spent \emph{more time to build and train their model reach inferior results} compared to methods that finish training more quickly. We primarily attribute this discrepancy between methods to the number of plans considered.

For example, during the training on JOB, Neo executed between 4,000 and 8,000 plans in PostgreSQL, Balsa between 19,000 and 21,000 plans. Even ignoring the quality of either methods' executed plans, it is obvious that 2-3$\times$ more plans also require more processing time. The authors of Balsa specifically tackled this challenge by allowing all required plans to be executed on multiple DBMS instances in parallel and by timing out long-running queries, which Neo does not. LEON does not fully execute the majority of its generated plans; However, it calls PostgreSQL to ask for cost estimates up to multiple tens of thousands of subplans, such that predicting just a plan for query 29a (the query with 17 aliased tables, the highest amount in all of JOB) takes around 6.5 hours\footnote{LEON caches plan and subplan cost estimates, generating files on the hard disk as large as 1.7 GB for JOB and 120 MB for STACK, respectively.}.


\subsection{Ablation Study: Covariate Shift}
\label{ablation_study__covariate_shift}

One of the challenges for query optimizers, in general, is their dependency on up-to-date statistics of the database content. In DBMS, statistics are regularly refreshed, but LQOs do not have the luxury to easily update trained model weights, with options to either train a new model from scratch or fine-tune and continue training, adapting to changes to the underlying database.

To show whether an encoding that represents the content of the database solely by cardinality (such as Bao) can deal with covariate shift, we conduct the following experiment. We generate a smaller copy of IMDB, referred to as IMDB-50\%. As the name implies, we keep 50\% of the rows in the \texttt{title} table using Bernoulli sampling, ensuring that the available data is halved, but the distribution of values remains comparable to the original version. The other 50\% of rows are dropped using CASCADE to ensure referential integrity. We specifically choose to alter the contents of the \texttt{title} table since it is the only table in IMDB that is part of all JOB queries. 

After sampling, we see a reduction of 50\% for the number of records in all movie-related (title, movie\_companies, movie\_info, movie\_info\_idx, movie\_keyword, and movie\_link) and cast-related tables (cast\_info and complete\_cast). Our sampling on \texttt{title} leaves all other tables unaffected. After the changes had been made to IMDB-50\%, the internal statistics of PostgreSQL were updated.


Our experiment aims to show that methods like Bao only using the cardinality in their encoding show a performance degradation when more data is added (simulating covariate shift). We train one Bao model on IMDB (referred to as Bao-Full) and a second Bao model on the reduced size IMDB-50\% (referred to as Bao-50) using the same "base query split 1" train/test split. 

Query 16b is a striking outlier, timing out in 1 of 4 Bao-50 models, while the other 3 generated plans that are 19 seconds slower than Bao-Full. In relative differences, query 31c is 24$\times$ slower using Bao-50 at 8.4 seconds compared to Bao-Full with 350 milliseconds. Query 17a is 4.5$\times$ slower at 12.2 seconds compared to Bao-Full with 2.7 seconds. On the other hand, the different cardinality regimes seen by Bao-50 also allow it to improve a few queries over Bao-Full by a factor of 1.9$\times$ for query 7c, 1.6$\times$ for 26c and 1.3$\times$ for 10c.

These results indicate that the \emph{DBMS system updating the statistics (i.e., cardinality estimates) is insufficient to keep up with a newly trained model}. This performance degradation further points to difficulties in generalization, particularly when larger cardinality values have not been seen during the training process and are, hence, out of distribution. There is currently no solution to this problem other than re-training or fine-tuning the model with queries running against the new database state. Because of this, methods that can continuously be updated and re-trained are preferable.


\subsection{Ablation Study: Bitmap and Tid Scans}
\label{sec:bitmap_tid_experiment}

We have observed multiple publications that disabled bitmap and tid scans, namely Balsa \cite{balsa_Yang2022}, LEON \cite{chen2023leon}, and a recently published analysis \cite{zhang2023adaptive}, without giving a reason for doing so. This experiment aims to see if changing PostgreSQL's tool kit significantly impacts the query performance of the individual queries. For the comparison, we use the baseline PostgreSQL performance from the previous experiment in Section \ref{subsec:comparison_sota}, and we have run the same 113 queries from JOB with bitmap and tid scans disabled.

The difference in execution time exceeds 250 milliseconds for 28 queries, 24 of which are statistically significant.
For those 24 queries, disabling bitmap and tid scans \textbf{speeds up} queries 28a, 7c, and 30a relative to their original execution times by a factor of $5.5\times$, $2.0\times$, and $1.8\times$, respectively. In contrast, queries 30c, 28b, and 15c are \textbf{slowed down} by a factor of $2.4\times$, $1.9\times$ and $1.5\times$, respectively.

These findings show that \emph{allowing PostgreSQL to use bitmap and tid scans significantly impacts the query performance}, particularly for the query templates 7, 8, 28, and 30. An interesting observation here is that the same family that has the highest gain of disabling said scans (query 28a with a speedup of $5.5\times$) also features a large slowdown (query 28b with a slow down by $1.9\times$). 


\subsection{Ablation Study: Genetic Query Optimizer}
\label{sec:geqo_experiment}

Similar to the disabling of various scan types, there exist differences in using GEQO, i.e., PostgreSQL's genetic query optimizer, across recent publications. We have analyzed the impact of disabling GEQO on the execution time of queries from JOB. Our experiment revealed 5 queries, for which the difference is statistically significant.
Disabling GEQO \textbf{speeds up} query 30a by a factor of $1.6\times$, while the other four queries are \textbf{slowed down} by a factor of $9.9\times$ (24b), $2.2\times$ (26c), $2.1\times$ (28a) and $1.7\times$ (28b). The large slow-down factor of query 24b is explained by its quick execution time of just 28 milliseconds versus 272 milliseconds with GEQO disabled. While the impact of GEQO is smaller than that of bitmap and tid scans, it remains significant in particular among the slowest query templates.

In summary, these results show that it is \emph{paramount that PostgreSQL operates at full capacity} (i.e., with GEQO enabled) in particular when the LQO does not replace, but rather enhance or guide the existing optimizer (for example, through the use of hints).

\subsection{Robustness of Query Execution Times}
\label{sec:analysis_of_successive_query_execution_times}

The goal of this experiment is to determine the choice of the number of repeated executions $k$ for the same query to reach a consistent execution time in a hot cache scenario. For example, in LEON~\cite{chen2023leon} the authors use the geometric mean with $k=3$, while in \cite{zhang2023adaptive} the authors execute queries $k=5$ times and take the arithmetic mean.

To achieve a fair comparison, we executed all queries of JOB using \texttt{EXPLAIN} \texttt{ANALYZE} 50 times in succession and in order (i.e., 1a, 1a, 1a, ..., 1a, 1b, 1b, ...). The execution time is extracted from PostgreSQL's \texttt{EXPLAIN} response, removing the network latency to the database from our measurements. By evaluating the distribution of execution times for the $k$-th iteration empirically, we can propose a value of $k$ that strikes the balance between costs and robustness.

\begin{figure}[h!]
    \centering
    \includegraphics[width=0.8\linewidth, keepaspectratio]{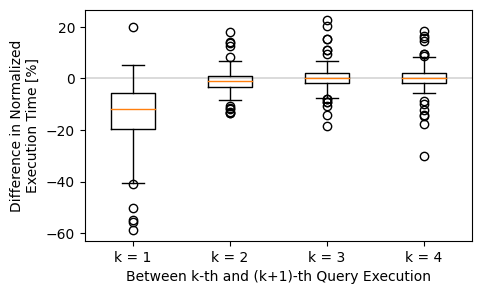}
    \caption{Difference in normalized execution time between successive query executions. For example, k=1 shows the difference between the 1st and 2nd query execution.}
    \label{fig:cold_warm_cache_experiment}
\end{figure}

Figure \ref{fig:cold_warm_cache_experiment} shows the normalized difference in query execution time (relative difference to the first executed query) when comparing pairs of the $k$-th and $(k+1)$-th query execution. We observe, that the query execution time significantly shifts for the majority of executed queries at $k=1$, with a mean reduction of 14.6\% between the 1st and 2nd query execution, and another 1.03\% from the 2nd to the 3rd. From then on, the fluctuations no longer show a trend that would benefit from more executions.

We see, thus empirically, that for robust measurements of execution times, it is important to at least execute a query twice. If time and costs allow, a third execution further improves the robustness, after which one can safely stop. Compared to the choice of $k=5$ in \cite{zhang2023adaptive}, taking the third execution is 40\% faster and more robust than averaging three measurements (for which the first query execution typically dominates as an outlier measurement).

\subsection{Analysis of Query Plan Types}
\label{sec:analysis_of_query_plans}

Given that there exists a larger number of bushy compared to left-deep and right-deep plans\footnote{Left-deep and right-deep plans are hereafter only referred to as left-deep plans, without loss of generality.}, it needs to be asked whether omitting \textit{bushy plans} (as for example in RTOS, LOGER and HybridQO) is a reasonable choice. In \cite{JOB_leis2015good}, the slowdown for restricted tree shapes was measured in comparison to the optimal plan.
The experiments' outcome shows that \emph{left-deep trees are worse than bushy ones} but still result in a reasonable performance. It is worth noting that these experiments were executed by injecting \textit{true cardinalities} to the cost model of the query optimizer. Moreover, some constraints on the join method selection according to \cite{garcia_molina2008dbsystems} were applied.

By forcing all combinations, we analyzed all possible plans for JOB queries with $\le5$ joins in the spirit of \cite{JOB_leis2015good}. However, rather than using true cardinalities (which are considered the optimal case), we ran our experiments with the \textit{DBMS's internal cardinality estimator}. Moreover, we allowed all join methods to be used. 

As a result, \emph{bushy plans perform on average like left-deep plans}. We confirm our results by obtaining the minimum of \emph{p-value=0.285} for a two-side Mann-Whitney U-test~\cite{mannwhitney1947utest}\footnote{The selection of the non-parametric test over the T-test stems from the observed lack of normal distribution plausibility across distinct logical and physical plans.} for the means of execution times.
At the left tail of the combined distribution of execution times (i.e. among the fastest plans), however, bushy trees are significantly superior with a \emph{p-value of 0.015} for the alternative hypothesis. That means, removing bushy trees from consideration drastically lowers the chance that a model finds the best plan.


\section{Conclusion}
\label{sec:conclusion}

In this paper, we outline the limitations of current LQO methods and put an emphasis on previously under-reported challenges. We provide a framework to equalize many parameters involved in benchmarking to yield increasingly robust results. 

We perform an evaluation of current LQO methods on the Join Order Benchmark and show that consistently outperforming PostgreSQL is more difficult than expected, particularly when looking at the query optimization problem as an end-to-end process. We believe that our paper is a first step towards reproducible and consistent benchmark evaluations for LQOs and thus provides important novel insights into LQOs from an ML perspective.

\begin{acks}

The project has received funding from the Swiss National Science Foundation under grant number 1921052. We also thank our colleagues from University of Konstanz, namely Michael Grossniklaus, Mehmet Aytimur and Silvan Reiner, and Dennis Gehrig from Zurich University of Applied Sciences, for valuable discussions.
\end{acks}

\vfill\null
\pagebreak

\bibliographystyle{ACM-Reference-Format}
\bibliography{main}


\begin{thebibliography}{44}


\ifx \showCODEN    \undefined \def \showCODEN     #1{\unskip}     \fi
\ifx \showDOI      \undefined \def \showDOI       #1{#1}\fi
\ifx \showISBNx    \undefined \def \showISBNx     #1{\unskip}     \fi
\ifx \showISBNxiii \undefined \def \showISBNxiii  #1{\unskip}     \fi
\ifx \showISSN     \undefined \def \showISSN      #1{\unskip}     \fi
\ifx \showLCCN     \undefined \def \showLCCN      #1{\unskip}     \fi
\ifx \shownote     \undefined \def \shownote      #1{#1}          \fi
\ifx \showarticletitle \undefined \def \showarticletitle #1{#1}   \fi
\ifx \showURL      \undefined \def \showURL       {\relax}        \fi
\providecommand\bibfield[2]{#2}
\providecommand\bibinfo[2]{#2}
\providecommand\natexlab[1]{#1}
\providecommand\showeprint[2][]{arXiv:#2}

\bibitem[\protect\citeauthoryear{Avnur and Hellerstein}{Avnur and
  Hellerstein}{2000}]%
        {hellerstein2000dynamic}
\bibfield{author}{\bibinfo{person}{Ron Avnur} {and} \bibinfo{person}{Joseph~M.
  Hellerstein}.} \bibinfo{year}{2000}\natexlab{}.
\newblock \showarticletitle{Eddies: Continuously Adaptive Query Processing}.
\newblock \bibinfo{journal}{\emph{SIGMOD Rec.}} \bibinfo{volume}{29},
  \bibinfo{number}{2} (\bibinfo{date}{may} \bibinfo{year}{2000}),
  \bibinfo{pages}{261–272}.
\newblock
\showISSN{0163-5808}
\urldef\tempurl%
\url{https://doi.org/10.1145/335191.335420}
\showDOI{\tempurl}


\bibitem[\protect\citeauthoryear{Brownlee}{Brownlee}{2020}]%
        {brownlee2020data}
\bibfield{author}{\bibinfo{person}{Jason Brownlee}.}
  \bibinfo{year}{2020}\natexlab{}.
\newblock \bibinfo{booktitle}{\emph{Data preparation for machine learning: data
  cleaning, feature selection, and data transforms in Python}}.
\newblock \bibinfo{publisher}{Machine Learning Mastery}.
\newblock


\bibitem[\protect\citeauthoryear{Chen, Gao, Chen, and Tu}{Chen
  et~al\mbox{.}}{2023b}]%
        {chen2023loger}
\bibfield{author}{\bibinfo{person}{Tianyi Chen}, \bibinfo{person}{Jun Gao},
  \bibinfo{person}{Hedui Chen}, {and} \bibinfo{person}{Yaofeng Tu}.}
  \bibinfo{year}{2023}\natexlab{b}.
\newblock \showarticletitle{LOGER: A Learned Optimizer Towards Generating
  Efficient and Robust Query Execution Plans}.
\newblock \bibinfo{journal}{\emph{Proceedings of the VLDB Endowment}}
  \bibinfo{volume}{16}, \bibinfo{number}{7} (\bibinfo{year}{2023}),
  \bibinfo{pages}{1777--1789}.
\newblock


\bibitem[\protect\citeauthoryear{Chen, Chen, Liang, Liu, Wang, Zeng, Su, and
  Zheng}{Chen et~al\mbox{.}}{2023a}]%
        {chen2023leon}
\bibfield{author}{\bibinfo{person}{Xu Chen}, \bibinfo{person}{Haitian Chen},
  \bibinfo{person}{Zibo Liang}, \bibinfo{person}{Shuncheng Liu},
  \bibinfo{person}{Jinghong Wang}, \bibinfo{person}{Kai Zeng},
  \bibinfo{person}{Han Su}, {and} \bibinfo{person}{Kai Zheng}.}
  \bibinfo{year}{2023}\natexlab{a}.
\newblock \showarticletitle{LEON: A New Framework for ML-Aided Query
  Optimization.}
\newblock \bibinfo{journal}{\emph{Proc. VLDB Endow.}} \bibinfo{volume}{16},
  \bibinfo{number}{9} (\bibinfo{year}{2023}), \bibinfo{pages}{2261--2273}.
\newblock


\bibitem[\protect\citeauthoryear{Dwivedi and Bresson}{Dwivedi and
  Bresson}{2021}]%
        {dwivedi2021gt}
\bibfield{author}{\bibinfo{person}{Vijay~Prakash Dwivedi} {and}
  \bibinfo{person}{Xavier Bresson}.} \bibinfo{year}{2021}\natexlab{}.
\newblock \bibinfo{title}{A Generalization of Transformer Networks to Graphs}.
\newblock
\newblock
\showeprint[arxiv]{2012.09699}~[cs.LG]


\bibitem[\protect\citeauthoryear{Garcia-Molina, Ullman, and
  Widom}{Garcia-Molina et~al\mbox{.}}{2008}]%
        {garcia_molina2008dbsystems}
\bibfield{author}{\bibinfo{person}{Hector Garcia-Molina},
  \bibinfo{person}{Jeffrey~D. Ullman}, {and} \bibinfo{person}{Jennifer Widom}.}
  \bibinfo{year}{2008}\natexlab{}.
\newblock \bibinfo{booktitle}{\emph{Database Systems: The Complete Book}
  (\bibinfo{edition}{2} ed.)}.
\newblock \bibinfo{publisher}{Prentice Hall Press}, \bibinfo{address}{USA}.
\newblock
\showISBNx{9780131873254}


\bibitem[\protect\citeauthoryear{Goodfellow, Bengio, and Courville}{Goodfellow
  et~al\mbox{.}}{2016}]%
        {goodfellow2016dl}
\bibfield{author}{\bibinfo{person}{Ian~J. Goodfellow}, \bibinfo{person}{Yoshua
  Bengio}, {and} \bibinfo{person}{Aaron Courville}.}
  \bibinfo{year}{2016}\natexlab{}.
\newblock \bibinfo{booktitle}{\emph{Deep Learning}}.
\newblock \bibinfo{publisher}{MIT Press}, \bibinfo{address}{Cambridge, MA,
  USA}.
\newblock
\newblock
\shownote{\url{http://www.deeplearningbook.org}.}


\bibitem[\protect\citeauthoryear{Guan, Li, Duan, Li, Ren, Sun, and Cheng}{Guan
  et~al\mbox{.}}{2021}]%
        {guan2021direct}
\bibfield{author}{\bibinfo{person}{Yang Guan}, \bibinfo{person}{Shengbo~Eben
  Li}, \bibinfo{person}{Jingliang Duan}, \bibinfo{person}{Jie Li},
  \bibinfo{person}{Yangang Ren}, \bibinfo{person}{Qi Sun}, {and}
  \bibinfo{person}{Bo Cheng}.} \bibinfo{year}{2021}\natexlab{}.
\newblock \bibinfo{title}{Direct and indirect reinforcement learning}.
\newblock
\newblock
\showeprint[arxiv]{1912.10600}~[cs.LG]


\bibitem[\protect\citeauthoryear{Guyon and Elisseeff}{Guyon and
  Elisseeff}{2006}]%
        {guyon2006features}
\bibfield{author}{\bibinfo{person}{Isabelle Guyon} {and}
  \bibinfo{person}{Andr{\'e} Elisseeff}.} \bibinfo{year}{2006}\natexlab{}.
\newblock \bibinfo{booktitle}{\emph{An Introduction to Feature Extraction}}.
\newblock \bibinfo{publisher}{Springer Berlin Heidelberg},
  \bibinfo{address}{Berlin, Heidelberg}, \bibinfo{pages}{1--25}.
\newblock
\showISBNx{978-3-540-35488-8}
\urldef\tempurl%
\url{https://doi.org/10.1007/978-3-540-35488-8_1}
\showDOI{\tempurl}


\bibitem[\protect\citeauthoryear{Han, Wu, Wu, Zhu, Yang, Liang, Zeng, Cong,
  Qin, Pfadler, Qian, Zhou, Li, and Cui}{Han et~al\mbox{.}}{2022}]%
        {han2021CEbenchmark}
\bibfield{author}{\bibinfo{person}{Yuxing Han}, \bibinfo{person}{Ziniu Wu},
  \bibinfo{person}{Peizhi Wu}, \bibinfo{person}{Rong Zhu},
  \bibinfo{person}{Jingyi Yang}, \bibinfo{person}{Tan~Wei Liang},
  \bibinfo{person}{Kai Zeng}, \bibinfo{person}{Gao Cong},
  \bibinfo{person}{Yanzhao Qin}, \bibinfo{person}{Andreas Pfadler},
  \bibinfo{person}{Zhengping Qian}, \bibinfo{person}{Jingren Zhou},
  \bibinfo{person}{Jiangneng Li}, {and} \bibinfo{person}{Bin Cui}.}
  \bibinfo{year}{2022}\natexlab{}.
\newblock \showarticletitle{Cardinality Estimation in DBMS: A Comprehensive
  Benchmark Evaluation}.
\newblock \bibinfo{journal}{\emph{VLDB}} \bibinfo{volume}{15},
  \bibinfo{number}{4} (\bibinfo{year}{2022}).
\newblock
\showISSN{2150-8097}


\bibitem[\protect\citeauthoryear{Heitz and Stockinger}{Heitz and
  Stockinger}{2019}]%
        {heitz2019join}
\bibfield{author}{\bibinfo{person}{Jonas Heitz} {and} \bibinfo{person}{Kurt
  Stockinger}.} \bibinfo{year}{2019}\natexlab{}.
\newblock \showarticletitle{Join query optimization with deep reinforcement
  learning algorithms}.
\newblock \bibinfo{journal}{\emph{arXiv preprint arXiv:1911.11689}}
  (\bibinfo{year}{2019}).
\newblock


\bibitem[\protect\citeauthoryear{Hilprecht, Schmidt, Kulessa, Molina, Kersting,
  and Binnig}{Hilprecht et~al\mbox{.}}{2020}]%
        {benjamin2020deepdb}
\bibfield{author}{\bibinfo{person}{Benjamin Hilprecht},
  \bibinfo{person}{Andreas Schmidt}, \bibinfo{person}{Moritz Kulessa},
  \bibinfo{person}{Alejandro Molina}, \bibinfo{person}{Kristian Kersting},
  {and} \bibinfo{person}{Carsten Binnig}.} \bibinfo{year}{2020}\natexlab{}.
\newblock \showarticletitle{DeepDB: Learn from Data, Not from Queries!}
\newblock \bibinfo{journal}{\emph{Proc. VLDB Endow.}} \bibinfo{volume}{13},
  \bibinfo{number}{7} (\bibinfo{date}{mar} \bibinfo{year}{2020}),
  \bibinfo{pages}{992–1005}.
\newblock
\showISSN{2150-8097}
\urldef\tempurl%
\url{https://doi.org/10.14778/3384345.3384349}
\showDOI{\tempurl}


\bibitem[\protect\citeauthoryear{Kacimi and Neumann}{Kacimi and
  Neumann}{2009}]%
        {kacimi2009systemr}
\bibfield{author}{\bibinfo{person}{Mouna Kacimi} {and} \bibinfo{person}{Thomas
  Neumann}.} \bibinfo{year}{2009}\natexlab{}.
\newblock \bibinfo{booktitle}{\emph{System R (R*) Optimizer}}.
\newblock \bibinfo{publisher}{Springer US}, \bibinfo{address}{Boston, MA},
  \bibinfo{pages}{2900--2905}.
\newblock
\showISBNx{978-0-387-39940-9}
\urldef\tempurl%
\url{https://doi.org/10.1007/978-0-387-39940-9_384}
\showDOI{\tempurl}


\bibitem[\protect\citeauthoryear{Kipf and Welling}{Kipf and Welling}{2017}]%
        {kipf2016gcnns}
\bibfield{author}{\bibinfo{person}{Thomas~N. Kipf} {and} \bibinfo{person}{Max
  Welling}.} \bibinfo{year}{2017}\natexlab{}.
\newblock \showarticletitle{{Semi-Supervised Classification with Graph
  Convolutional Networks}}. In \bibinfo{booktitle}{\emph{Proceedings of the 5th
  International Conference on Learning Representations}} (Palais des
  Congr{\`e}s Neptune, Toulon, France) \emph{(\bibinfo{series}{ICLR '17})}.
\newblock
\urldef\tempurl%
\url{https://openreview.net/forum?id=SJU4ayYgl}
\showURL{%
\tempurl}


\bibitem[\protect\citeauthoryear{Krishnan, Yang, Goldberg, Hellerstein, and
  Stoica}{Krishnan et~al\mbox{.}}{2018a}]%
        {krishnan2018dq}
\bibfield{author}{\bibinfo{person}{Sanjay Krishnan}, \bibinfo{person}{Zongheng
  Yang}, \bibinfo{person}{Kenneth Goldberg}, \bibinfo{person}{Joseph
  Hellerstein}, {and} \bibinfo{person}{Ion Stoica}.}
  \bibinfo{year}{2018}\natexlab{a}.
\newblock \showarticletitle{Learning to Optimize Join Queries With Deep
  Reinforcement Learning}.
\newblock  (\bibinfo{date}{08} \bibinfo{year}{2018}).
\newblock


\bibitem[\protect\citeauthoryear{Krishnan, Yang, Goldberg, Hellerstein, and
  Stoica}{Krishnan et~al\mbox{.}}{2018b}]%
        {krishnan2018learning}
\bibfield{author}{\bibinfo{person}{Sanjay Krishnan}, \bibinfo{person}{Zongheng
  Yang}, \bibinfo{person}{Ken Goldberg}, \bibinfo{person}{Joseph Hellerstein},
  {and} \bibinfo{person}{Ion Stoica}.} \bibinfo{year}{2018}\natexlab{b}.
\newblock \showarticletitle{Learning to optimize join queries with deep
  reinforcement learning}.
\newblock \bibinfo{journal}{\emph{arXiv preprint arXiv:1808.03196}}
  (\bibinfo{year}{2018}).
\newblock


\bibitem[\protect\citeauthoryear{La~Salle}{La~Salle}{1976}]%
        {la1976stability}
\bibfield{author}{\bibinfo{person}{Joseph~P La~Salle}.}
  \bibinfo{year}{1976}\natexlab{}.
\newblock \bibinfo{booktitle}{\emph{The stability of dynamical systems}}.
\newblock \bibinfo{publisher}{SIAM}.
\newblock


\bibitem[\protect\citeauthoryear{Leis, Gubichev, Mirchev, Boncz, Kemper, and
  Neumann}{Leis et~al\mbox{.}}{2015}]%
        {JOB_leis2015good}
\bibfield{author}{\bibinfo{person}{Viktor Leis}, \bibinfo{person}{Andrey
  Gubichev}, \bibinfo{person}{Atanas Mirchev}, \bibinfo{person}{Peter Boncz},
  \bibinfo{person}{Alfons Kemper}, {and} \bibinfo{person}{Thomas Neumann}.}
  \bibinfo{year}{2015}\natexlab{}.
\newblock \showarticletitle{How good are query optimizers, really?}
\newblock \bibinfo{journal}{\emph{Proceedings of the VLDB Endowment}}
  \bibinfo{volume}{9}, \bibinfo{number}{3} (\bibinfo{year}{2015}),
  \bibinfo{pages}{204--215}.
\newblock


\bibitem[\protect\citeauthoryear{Levandoski, Larson, and Stoica}{Levandoski
  et~al\mbox{.}}{2013}]%
        {levandoski2013identifying}
\bibfield{author}{\bibinfo{person}{Justin~J Levandoski},
  \bibinfo{person}{Per-{\AA}ke Larson}, {and} \bibinfo{person}{Radu Stoica}.}
  \bibinfo{year}{2013}\natexlab{}.
\newblock \showarticletitle{Identifying hot and cold data in main-memory
  databases}. In \bibinfo{booktitle}{\emph{2013 IEEE 29th International
  Conference on Data Engineering (ICDE)}}. IEEE, \bibinfo{pages}{26--37}.
\newblock


\bibitem[\protect\citeauthoryear{Liu}{Liu}{2009}]%
        {liu2009ltr}
\bibfield{author}{\bibinfo{person}{Tie-Yan Liu}.}
  \bibinfo{year}{2009}\natexlab{}.
\newblock \showarticletitle{Learning to Rank for Information Retrieval}.
\newblock \bibinfo{journal}{\emph{Foundations and Trends® in Information
  Retrieval}} \bibinfo{volume}{3}, \bibinfo{number}{3} (\bibinfo{year}{2009}),
  \bibinfo{pages}{225--331}.
\newblock
\showISSN{1554-0669}
\urldef\tempurl%
\url{https://doi.org/10.1561/1500000016}
\showDOI{\tempurl}


\bibitem[\protect\citeauthoryear{Mann and Whitney}{Mann and Whitney}{1947}]%
        {mannwhitney1947utest}
\bibfield{author}{\bibinfo{person}{H.~B. Mann} {and} \bibinfo{person}{D.~R.
  Whitney}.} \bibinfo{year}{1947}\natexlab{}.
\newblock \showarticletitle{On a Test of Whether one of Two Random Variables is
  Stochastically Larger than the Other}.
\newblock \bibinfo{journal}{\emph{The Annals of Mathematical Statistics}}
  \bibinfo{volume}{18}, \bibinfo{number}{1} (\bibinfo{year}{1947}),
  \bibinfo{pages}{50--60}.
\newblock
\showISSN{00034851}
\urldef\tempurl%
\url{http://www.jstor.org/stable/2236101}
\showURL{%
\tempurl}


\bibitem[\protect\citeauthoryear{Marcus, Negi, Mao, Tatbul, Alizadeh, and
  Kraska}{Marcus et~al\mbox{.}}{2021}]%
        {marcus2021bao}
\bibfield{author}{\bibinfo{person}{Ryan Marcus}, \bibinfo{person}{Parimarjan
  Negi}, \bibinfo{person}{Hongzi Mao}, \bibinfo{person}{Nesime Tatbul},
  \bibinfo{person}{Mohammad Alizadeh}, {and} \bibinfo{person}{Tim Kraska}.}
  \bibinfo{year}{2021}\natexlab{}.
\newblock \showarticletitle{Bao: Making learned query optimization practical}.
  In \bibinfo{booktitle}{\emph{Proceedings of the 2021 International Conference
  on Management of Data}}. \bibinfo{pages}{1275--1288}.
\newblock


\bibitem[\protect\citeauthoryear{Marcus, Negi, Mao, Zhang, Alizadeh, Kraska,
  Papaemmanouil, and Tatbul}{Marcus et~al\mbox{.}}{2019}]%
        {marcus12neo}
\bibfield{author}{\bibinfo{person}{Ryan Marcus}, \bibinfo{person}{Parimarjan
  Negi}, \bibinfo{person}{Hongzi Mao}, \bibinfo{person}{Chi Zhang},
  \bibinfo{person}{Mohammad Alizadeh}, \bibinfo{person}{Tim Kraska},
  \bibinfo{person}{Olga Papaemmanouil}, {and} \bibinfo{person}{Nesime Tatbul}.}
  \bibinfo{year}{2019}\natexlab{}.
\newblock \showarticletitle{Neo: A Learned Query Optimizer}.
\newblock \bibinfo{journal}{\emph{Proceedings of the VLDB Endowment}}
  \bibinfo{volume}{12}, \bibinfo{number}{11} (\bibinfo{year}{2019}).
\newblock


\bibitem[\protect\citeauthoryear{Marcus and Papaemmanouil}{Marcus and
  Papaemmanouil}{2018a}]%
        {marcus2018rejoin2}
\bibfield{author}{\bibinfo{person}{Ryan Marcus} {and} \bibinfo{person}{Olga
  Papaemmanouil}.} \bibinfo{year}{2018}\natexlab{a}.
\newblock \showarticletitle{Deep Reinforcement Learning for Join Order
  Enumeration}. In \bibinfo{booktitle}{\emph{Proceedings of the First
  International Workshop on Exploiting Artificial Intelligence Techniques for
  Data Management}} (Houston, TX, USA) \emph{(\bibinfo{series}{aiDM'18})}.
  \bibinfo{publisher}{Association for Computing Machinery},
  \bibinfo{address}{New York, NY, USA}, Article \bibinfo{articleno}{3},
  \bibinfo{numpages}{4}~pages.
\newblock
\showISBNx{9781450358514}
\urldef\tempurl%
\url{https://doi.org/10.1145/3211954.3211957}
\showDOI{\tempurl}


\bibitem[\protect\citeauthoryear{Marcus and Papaemmanouil}{Marcus and
  Papaemmanouil}{2018b}]%
        {marcus2018rejoin1}
\bibfield{author}{\bibinfo{person}{Ryan Marcus} {and} \bibinfo{person}{Olga
  Papaemmanouil}.} \bibinfo{year}{2018}\natexlab{b}.
\newblock \showarticletitle{Towards a Hands-Free Query Optimizer through Deep
  Learning}.
\newblock  (\bibinfo{date}{09} \bibinfo{year}{2018}).
\newblock


\bibitem[\protect\citeauthoryear{Mou, Li, Zhang, Wang, and Jin}{Mou
  et~al\mbox{.}}{2016}]%
        {mou2016treeconv}
\bibfield{author}{\bibinfo{person}{Lili Mou}, \bibinfo{person}{Ge Li},
  \bibinfo{person}{Lu Zhang}, \bibinfo{person}{Tao Wang}, {and}
  \bibinfo{person}{Zhi Jin}.} \bibinfo{year}{2016}\natexlab{}.
\newblock \showarticletitle{Convolutional Neural Networks over Tree Structures
  for Programming Language Processing}. In
  \bibinfo{booktitle}{\emph{Proceedings of the Thirtieth AAAI Conference on
  Artificial Intelligence}} (Phoenix, Arizona)
  \emph{(\bibinfo{series}{AAAI'16})}. \bibinfo{publisher}{AAAI Press},
  \bibinfo{pages}{1287–1293}.
\newblock


\bibitem[\protect\citeauthoryear{{Nippon Telegraph and Telephone
  Corporation}}{{Nippon Telegraph and Telephone Corporation}}{2012}]%
        {pg_hint_plan_doc}
\bibfield{author}{\bibinfo{person}{{Nippon Telegraph and Telephone
  Corporation}}.} \bibinfo{year}{2012}\natexlab{}.
\newblock \bibinfo{title}{{pg\_hint\_plan Documentation}}.
\newblock
  \bibinfo{howpublished}{\url{https://pghintplan.osdn.jp/pg_hint_plan.html}}.
\newblock
\newblock
\shownote{[Online; accessed August, 2023].}


\bibitem[\protect\citeauthoryear{Petkovi{\'{c}}}{Petkovi{\'{c}}}{2011}]%
        {petkovic2011dp_vs_ga}
\bibfield{author}{\bibinfo{person}{Du{\v{s}}an Petkovi{\'{c}}}.}
  \bibinfo{year}{2011}\natexlab{}.
\newblock \showarticletitle{Dynamic Programming Algorithm vs. Genetic
  Algorithm: Which is Faster?}. In \bibinfo{booktitle}{\emph{Research and
  Development in Intelligent Systems XXVII}},
  \bibfield{editor}{\bibinfo{person}{Max Bramer}, \bibinfo{person}{Miltos
  Petridis}, {and} \bibinfo{person}{Adrian Hopgood}} (Eds.).
  \bibinfo{publisher}{Springer London}, \bibinfo{address}{London},
  \bibinfo{pages}{483--488}.
\newblock
\showISBNx{978-0-85729-130-1}


\bibitem[\protect\citeauthoryear{Rogov}{Rogov}{2022}]%
        {rogov2022joinmethods}
\bibfield{author}{\bibinfo{person}{Egor Rogov}.}
  \bibinfo{year}{2022}\natexlab{}.
\newblock \bibinfo{title}{{Queries in PostgreSQL: Sort and Merge}}.
\newblock
  \bibinfo{howpublished}{\url{https://postgrespro.com/blog/pgsql/5969770}}.
\newblock
\newblock
\shownote{[Online; accessed August, 2023].}


\bibitem[\protect\citeauthoryear{Russell and Norvig}{Russell and
  Norvig}{2010}]%
        {russel2010ml}
\bibfield{author}{\bibinfo{person}{Stuart Russell} {and} \bibinfo{person}{Peter
  Norvig}.} \bibinfo{year}{2010}\natexlab{}.
\newblock \bibinfo{booktitle}{\emph{Artificial Intelligence: A Modern Approach}
  (\bibinfo{edition}{3} ed.)}.
\newblock \bibinfo{publisher}{Prentice Hall}.
\newblock


\bibitem[\protect\citeauthoryear{Smith, Treat, and Browne}{Smith
  et~al\mbox{.}}{2021}]%
        {pg_best_practices}
\bibfield{author}{\bibinfo{person}{Greg Smith}, \bibinfo{person}{Robert Treat},
  {and} \bibinfo{person}{Christopher Browne}.} \bibinfo{year}{2021}\natexlab{}.
\newblock \bibinfo{title}{{Tuning Your PostgreSQL Server}}.
\newblock
  \bibinfo{howpublished}{\url{https://wiki.postgresql.org/wiki/Tuning_Your_PostgreSQL_Server}}.
\newblock
\newblock
\shownote{[Online; accessed August, 2023].}


\bibitem[\protect\citeauthoryear{Sutton and Barto}{Sutton and Barto}{2018}]%
        {sutton1998rl}
\bibfield{author}{\bibinfo{person}{Richard~S. Sutton} {and}
  \bibinfo{person}{Andrew~G. Barto}.} \bibinfo{year}{2018}\natexlab{}.
\newblock \bibinfo{booktitle}{\emph{Reinforcement Learning: An Introduction}
  (\bibinfo{edition}{second} ed.)}.
\newblock \bibinfo{publisher}{The MIT Press}.
\newblock
\urldef\tempurl%
\url{http://incompleteideas.net/book/the-book-2nd.html}
\showURL{%
\tempurl}


\bibitem[\protect\citeauthoryear{Tai, Socher, and Manning}{Tai
  et~al\mbox{.}}{2015}]%
        {tai2015treelstm}
\bibfield{author}{\bibinfo{person}{Kai~Sheng Tai}, \bibinfo{person}{Richard
  Socher}, {and} \bibinfo{person}{Christopher~D. Manning}.}
  \bibinfo{year}{2015}\natexlab{}.
\newblock \showarticletitle{Improved Semantic Representations From
  Tree-Structured Long Short-Term Memory Networks}. In
  \bibinfo{booktitle}{\emph{Proceedings of the 53rd Annual Meeting of the
  Association for Computational Linguistics and the 7th International Joint
  Conference on Natural Language Processing (Volume 1: Long Papers)}}.
  \bibinfo{publisher}{Association for Computational Linguistics},
  \bibinfo{address}{Beijing, China}, \bibinfo{pages}{1556--1566}.
\newblock
\urldef\tempurl%
\url{https://doi.org/10.3115/v1/P15-1150}
\showDOI{\tempurl}


\bibitem[\protect\citeauthoryear{{The PostgreSQL Global Development
  Group}}{{The PostgreSQL Global Development Group}}{2023}]%
        {geqo_pg_doc}
\bibfield{author}{\bibinfo{person}{{The PostgreSQL Global Development Group}}.}
  \bibinfo{year}{2023}\natexlab{}.
\newblock \bibinfo{title}{{Genetic Query Optimization (GEQO) in PostgreSQL}}.
\newblock
  \bibinfo{howpublished}{\url{https://www.postgresql.org/docs/current/geqo-pg-intro.html}}.
\newblock
\newblock
\shownote{[Online; accessed August, 2023].}


\bibitem[\protect\citeauthoryear{{Transaction Processing Performance
  Council}}{{Transaction Processing Performance Council}}{2023}]%
        {tpc_benchmark_doc}
\bibfield{author}{\bibinfo{person}{{Transaction Processing Performance
  Council}}.} \bibinfo{year}{2023}\natexlab{}.
\newblock \bibinfo{title}{{TPC Benchmarks Overview}}.
\newblock
  \bibinfo{howpublished}{\url{https://www.tpc.org/information/benchmarks5.asp}}.
\newblock
\newblock
\shownote{[Online; accessed August, 2023].}


\bibitem[\protect\citeauthoryear{Wan, Zeiler, Zhang, Le~Cun, and Fergus}{Wan
  et~al\mbox{.}}{2013}]%
        {wan2013dropout}
\bibfield{author}{\bibinfo{person}{Li Wan}, \bibinfo{person}{Matthew Zeiler},
  \bibinfo{person}{Sixin Zhang}, \bibinfo{person}{Yann Le~Cun}, {and}
  \bibinfo{person}{Rob Fergus}.} \bibinfo{year}{2013}\natexlab{}.
\newblock \showarticletitle{Regularization of Neural Networks using
  DropConnect}. In \bibinfo{booktitle}{\emph{Proceedings of the 30th
  International Conference on Machine Learning}}
  \emph{(\bibinfo{series}{Proceedings of Machine Learning Research})},
  \bibfield{editor}{\bibinfo{person}{Sanjoy Dasgupta} {and}
  \bibinfo{person}{David McAllester}} (Eds.), Vol.~\bibinfo{volume}{28}.
  \bibinfo{publisher}{PMLR}, \bibinfo{address}{Atlanta, Georgia, USA},
  \bibinfo{pages}{1058--1066}.
\newblock
\urldef\tempurl%
\url{https://proceedings.mlr.press/v28/wan13.html}
\showURL{%
\tempurl}


\bibitem[\protect\citeauthoryear{Wang and Chen}{Wang and Chen}{1996}]%
        {wang1996complexity}
\bibfield{author}{\bibinfo{person}{Chihping Wang} {and}
  \bibinfo{person}{Ming-Syan Chen}.} \bibinfo{year}{1996}\natexlab{}.
\newblock \showarticletitle{On the complexity of distributed query
  optimization}.
\newblock \bibinfo{journal}{\emph{IEEE Transactions on Knowledge and Data
  Engineering}} \bibinfo{volume}{8}, \bibinfo{number}{4}
  (\bibinfo{year}{1996}), \bibinfo{pages}{650--662}.
\newblock


\bibitem[\protect\citeauthoryear{Webb}{Webb}{2010}]%
        {webb2010ml}
\bibfield{author}{\bibinfo{person}{Geoffrey~I. Webb}.}
  \bibinfo{year}{2010}\natexlab{}.
\newblock \bibinfo{booktitle}{\emph{Overfitting}}.
\newblock \bibinfo{publisher}{Springer US}, \bibinfo{address}{Boston, MA},
  \bibinfo{pages}{744--744}.
\newblock
\showISBNx{978-0-387-30164-8}
\urldef\tempurl%
\url{https://doi.org/10.1007/978-0-387-30164-8_623}
\showDOI{\tempurl}


\bibitem[\protect\citeauthoryear{Yang, Chiang, Luan, Mittal, Luo, and
  Stoica}{Yang et~al\mbox{.}}{2022}]%
        {balsa_Yang2022}
\bibfield{author}{\bibinfo{person}{Zongheng Yang}, \bibinfo{person}{Wei~Lin
  Chiang}, \bibinfo{person}{Sifei Luan}, \bibinfo{person}{Gautam Mittal},
  \bibinfo{person}{Michael Luo}, {and} \bibinfo{person}{Ion Stoica}.}
  \bibinfo{year}{2022}\natexlab{}.
\newblock \showarticletitle{Balsa: Learning a Query Optimizer Without Expert
  Demonstrations}.
\newblock \bibinfo{journal}{\emph{Proceedings of the ACM SIGMOD International
  Conference on Management of Data}} (\bibinfo{date}{6} \bibinfo{year}{2022}),
  \bibinfo{pages}{931--944}.
\newblock
\showISBNx{9781450392495}
\showISSN{07308078}
\urldef\tempurl%
\url{https://doi.org/10.1145/3514221.3517885}
\showDOI{\tempurl}


\bibitem[\protect\citeauthoryear{Yang, Kamsetty, Luan, Liang, Duan, Chen, and
  Stoica}{Yang et~al\mbox{.}}{2020}]%
        {yang2020neurocard}
\bibfield{author}{\bibinfo{person}{Zongheng Yang}, \bibinfo{person}{Amog
  Kamsetty}, \bibinfo{person}{Sifei Luan}, \bibinfo{person}{Eric Liang},
  \bibinfo{person}{Yan Duan}, \bibinfo{person}{Xi Chen}, {and}
  \bibinfo{person}{Ion Stoica}.} \bibinfo{year}{2020}\natexlab{}.
\newblock \showarticletitle{NeuroCard: One Cardinality Estimator for All
  Tables}.
\newblock \bibinfo{journal}{\emph{Proc. VLDB Endow.}} \bibinfo{volume}{14},
  \bibinfo{number}{1} (\bibinfo{date}{sep} \bibinfo{year}{2020}),
  \bibinfo{pages}{61–73}.
\newblock
\showISSN{2150-8097}
\urldef\tempurl%
\url{https://doi.org/10.14778/3421424.3421432}
\showDOI{\tempurl}


\bibitem[\protect\citeauthoryear{Yu, Chai, Li, and Liu}{Yu
  et~al\mbox{.}}{2022}]%
        {yu2022hybridqo}
\bibfield{author}{\bibinfo{person}{Xiang Yu}, \bibinfo{person}{Chengliang
  Chai}, \bibinfo{person}{Guoliang Li}, {and} \bibinfo{person}{Jiabin Liu}.}
  \bibinfo{year}{2022}\natexlab{}.
\newblock \showarticletitle{Cost-based or learning-based? A hybrid query
  optimizer for query plan selection}.
\newblock \bibinfo{journal}{\emph{Proceedings of the VLDB Endowment}}
  \bibinfo{volume}{15}, \bibinfo{number}{13} (\bibinfo{year}{2022}),
  \bibinfo{pages}{3924--3936}.
\newblock


\bibitem[\protect\citeauthoryear{Yu, Li, Chai, and Tang}{Yu
  et~al\mbox{.}}{2020}]%
        {yu2020rtos}
\bibfield{author}{\bibinfo{person}{Xiang Yu}, \bibinfo{person}{Guoliang Li},
  \bibinfo{person}{Chengliang Chai}, {and} \bibinfo{person}{Nan Tang}.}
  \bibinfo{year}{2020}\natexlab{}.
\newblock \showarticletitle{Reinforcement learning with tree-lstm for join
  order selection}. In \bibinfo{booktitle}{\emph{2020 IEEE 36th International
  Conference on Data Engineering (ICDE)}}. IEEE, \bibinfo{pages}{1297--1308}.
\newblock


\bibitem[\protect\citeauthoryear{Yunjia, Yannis, Jignesh~M., and
  Theodoros}{Yunjia et~al\mbox{.}}{2023}]%
        {zhang2023adaptive}
\bibfield{author}{\bibinfo{person}{Zhang Yunjia}, \bibinfo{person}{Chronis
  Yannis}, \bibinfo{person}{Patel Jignesh~M.}, {and}
  \bibinfo{person}{Rekatsinas Theodoros}.} \bibinfo{year}{2023}\natexlab{}.
\newblock \showarticletitle{Simple Adaptive Query Processing vs. Learned Query
  Optimizers: Observations and Analysis.}
\newblock \bibinfo{journal}{\emph{Proc. VLDB Endow.}} \bibinfo{volume}{16},
  \bibinfo{number}{9} (\bibinfo{year}{2023}), \bibinfo{pages}{2962--2975}.
\newblock


\bibitem[\protect\citeauthoryear{Zhu, Chen, Ding, Chen, Pfadler, Wu, and
  Zhou}{Zhu et~al\mbox{.}}{2023}]%
        {zhu2023lero}
\bibfield{author}{\bibinfo{person}{Rong Zhu}, \bibinfo{person}{Wei Chen},
  \bibinfo{person}{Bolin Ding}, \bibinfo{person}{Xingguang Chen},
  \bibinfo{person}{Andreas Pfadler}, \bibinfo{person}{Ziniu Wu}, {and}
  \bibinfo{person}{Jingren Zhou}.} \bibinfo{year}{2023}\natexlab{}.
\newblock \showarticletitle{Lero: A Learning-to-Rank Query Optimizer}.
\newblock \bibinfo{journal}{\emph{arXiv preprint arXiv:2302.06873}}
  (\bibinfo{year}{2023}).
\newblock


\end{thebibliography}

\appendix

\end{document}